\documentclass[a4paper,fleqn]{cas-dc}
\usepackage{graphicx} % Required for inserting images

\usepackage{array}
\usepackage{tabularx}
\usepackage[hypcap=false]{caption}

\usepackage{makecell}
\usepackage{url}

\usepackage{textcomp} 
\usepackage{float}
\usepackage{placeins}
\usepackage[numbers,sort&compress]{natbib}


\makeatletter
\def\fps@figure{!htbp}
\def\fps@table{!htbp}
\makeatother

\begin{document}

\title [mode = title]{An Eclipse-Ballooning Study of Shadow Bands During the April 2024 Total Eclipse}
\author[1]{Giana Deskevich}
\author[2]{Norris Bach}
\author[1]{Kristian Borysiak}
\author[1]{Russell J. Clark}
\author[1,3]{Louis W. Coban}
\author[1]{Istvan Danko}
\author[1]{Luke Docherty}
\author[1]{Michael Hatridge}
\author[1]{Howard Malc}
\author[1]{Boris Mestis}
\author[1]{Emma Moran}
\author[1]{Mathilda Nilsson}
\author[2]{Jeffrey B. Peterson}
\author[1,3]{Edward Michael Potosky}
\author[1]{Sandhya M. Rao}
\author[1]{Peri Schindelheim}
\author[4]{James D. Turnshek}
\author[1]{Ameya Velankar}
\author[1]{Ryan Young}
\author[1,3] {David A. Turnshek}

\affiliation[1]{organization={Department of Physics and Astronomy, University of Pittsburgh},
                city={Pittsburgh},
                postcode={PA 15260},
                country={USA}}
\affiliation[2]{organization={Department of Physics, Carnegie Mellon University},
                city={Pittsburgh},
                postcode={PA 15251},
                country={USA}}
\affiliation[3]{organization={Allegheny Observatory, 159 Riverview Ave, University of Pittsburgh},
                city={Pittsburgh},
                postcode={PA 15214},
                country={USA}}
\affiliation[4]{organization={Formant},
                city={San Francisco},
                postcode={CA 94105},
                country={USA}}
\shorttitle{Eclipse-Ballooning Study of Shadow Bands}
%\shortauthors{Deskevich et al.}

%\orcidauthor{0009-0002-2162-5918}{Giana Deskevich}

\begin{abstract}
In this study, we searched for shadow bands associated with the total solar eclipse of April 8, 2024, with the aim of improving our understanding of their origin. Shadow bands are debated to arise either from atmospheric turbulence within Earth’s planetary boundary layer (PBL) or from a diffraction-interference effect occurring above the atmosphere. To test these theories, high-altitude balloons (HABs) equipped with light sensors, ground-based light sensors, radiosondes launched with weather balloons, and an aircraft-mounted light sensor were deployed. Observations were conducted in Concan, TX, within the path of totality and from an aircraft operating over northeast Vermont to obtain clear-sky measurements. Unlike a 2017 high-altitude balloon study that reported a persistent 4.5~Hz signal attributed to shadow bands above the PBL and on the ground, no shadow band-like periodic photodiode signals were detected above the PBL in Texas or northeastern Vermont in the present study, despite the use of improved instrumentation. Cloud cover prevented useful ground-based measurements in Texas, limiting conclusions regarding shadow band behavior near the surface. These findings suggest that shadow bands may not always be present at all locations or, when present, may depend strongly on local atmospheric conditions. Together with earlier studies, these results highlight the need for continued multi-platform observations during future total solar eclipses.
\end{abstract}

\begin{keywords}
Total solar eclipse \sep shadow bands \sep atmospheric turbulence \sep planetary boundary layer \sep high-altitude balloons \sep radiosondes \sep spectrogram analysis \sep optical phenomena \sep light intensity measurements \sep diffraction-interference
\end{keywords}

\maketitle

\section{Introduction}
This paper presents an analysis of a search for shadow bands during the total solar eclipse of April 8, 2024. The investigation was part of a broader observational effort associated with recent solar eclipses.

Shadow bands are faint, rapidly moving patterns of light and dark that are sometimes observed on the ground, walls, or other bright surfaces near the beginning and end of totality during a solar eclipse. Nineteenth- and early twentieth-century reports describe them as wavy or undulatory bands, often compared to light reflected from rippling water, with reported widths and spacings ranging from a few centimeters to several decimeters \cite{ranyard1879, rotch1908}. The observations also show that shadow bands are not uniformly observed during every eclipse or at every totality location.

The physical origin of shadow bands remains uncertain. Historical record compilations emphasized their variable timing, motion, spacing, and apparent relationship to local observing conditions, while later interpretations commonly associated them with irregular refraction of the thin solar crescent through disturbed air \cite{rotch1908}. More recent studies have generally favored refractive distortions caused by atmospheric turbulence within the planetary boundary layer (PBL) \cite{codona1986, marschall1984}, although diffraction- or interference-based explanations have also been proposed historically and challenged on the basis of observed band speeds and spacings \cite{marschall1984, rotch1908}. Distinguishing between these possibilities requires observations made both within and above the turbulent boundary layer, alongside measurements of the atmospheric structure during totality.

Previous high-altitude eclipse observations provide the motivation for the present study. During the August 21, 2017 total solar eclipse, an eclipse-ballooning investigation reported a persistent 4.5 Hz signal in both ground-based and high-altitude balloon photodiode data and interpreted this signal as a possible shadow band signature detected above the lower turbulent atmosphere \cite{madhani2020}. Because shadow bands are traditionally associated with near-surface visual observations, a reported signal at high altitude raises the question of whether shadow band-like intensity fluctuations can occur above the PBL or whether such detections depend on local atmospheric and instrumental conditions.

To test this question during the April 8, 2024 total solar eclipse, photodiode light sensors were deployed onto two high-altitude balloons launched from Concan, Texas and on the ground at the same site. An additional set of light sensors were deployed on an aircraft operating above the lower atmosphere in northeastern Vermont. Radiosondes launched from Concan, together with radiosonde data from Pittsburg, New Hampshire, were used to characterize the height of the PBL and the atmospheric structure near totality. The light-sensor data were analyzed using spectrogram methods to search for periodic intensity fluctuations, and a signal-injection test was performed to evaluate whether shadow band-like signals would have been detectable in the 2024 data.

The primary objective of this study is to determine whether shadow band-like intensity fluctuations were present above the lower turbulent atmosphere during the April 8, 2024 eclipse. This test is motivated by the 2017 eclipse-ballooning result and by broader uncertainty surrounding the origin of shadow bands. In the 2024 data, no persistent frequency signature comparable to the 4.5 Hz signal reported in 2017 was detected in the high-altitude balloon or aircraft measurements. Because cloud cover made the Concan ground-based light-sensor data unusable for detecting near-surface shadow bands, the present observations cannot determine whether shadow bands were present near the surface at the site. Instead, the results constrain the presence of detectable shadow band-like signals above the PBL for the observing conditions sampled in this study.

The remainder of this paper is organized as follows. Section 2 describes the observing sites, instrumentation, and data collection, including the high-altitude balloon, aircraft, ground-based, and radiosonde measurements. Section 3 presents the data-processing methods, spectrogram analysis, signal-injection test, and radiosonde-based determination of the PBL height. Section 4 discusses the implications of the non-detections for the proposed shadow band mechanisms and summarizes the major findings. An Appendix includes additional results in the form of Figures.

\begin{figure*}
    \centering
    \includegraphics[width=\textwidth]{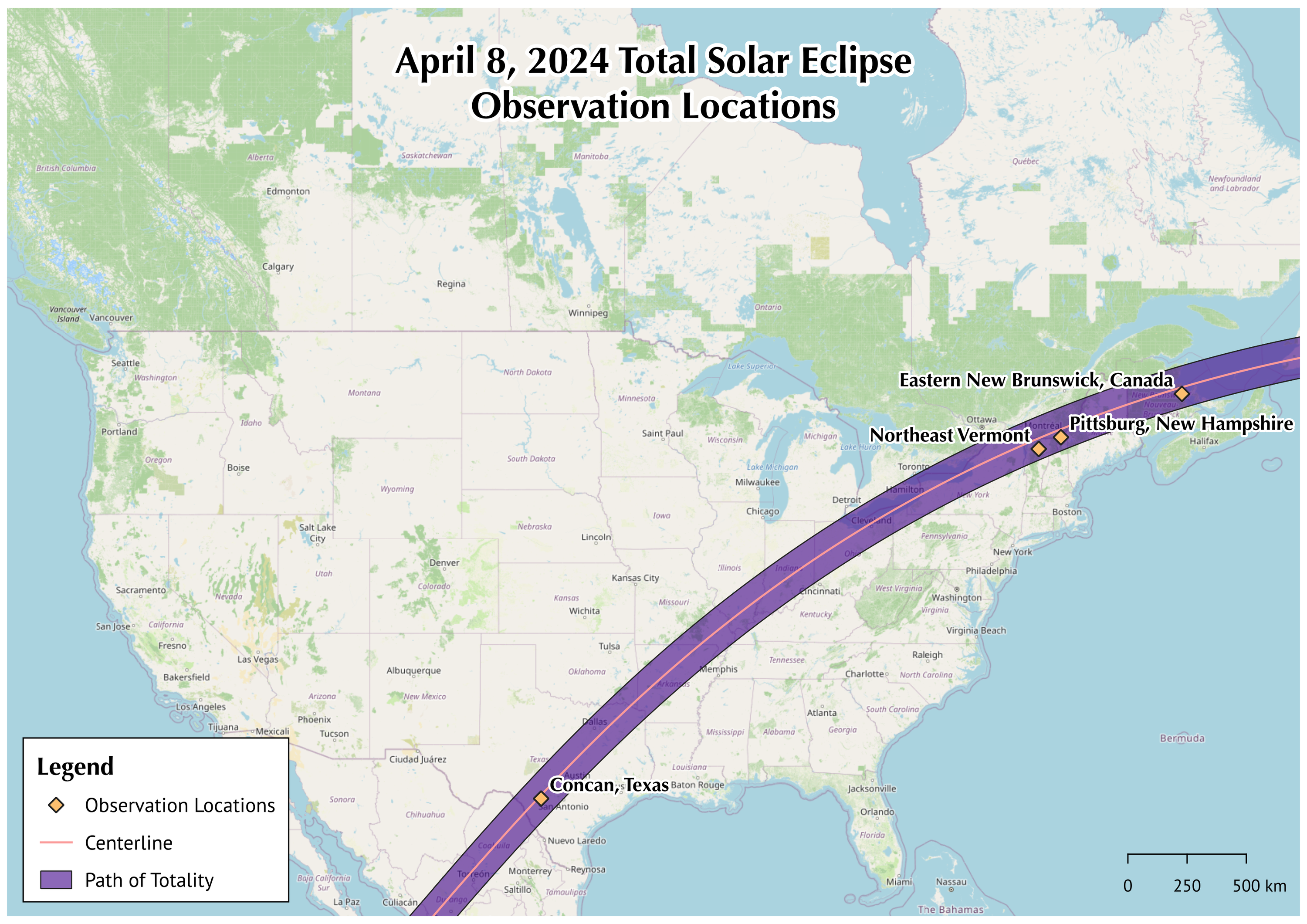}
    \caption{\textbf{Observation locations for the April 8, 2024 total solar eclipse.}
The map shows the eclipse path of totality, centerline, and the observing regions used in this study. The primary observing site was Concan, Texas ($29.488^\circ$N, $99.707^\circ$W), where HAB photodiodes, ground-based photodiodes, and radiosondes were deployed. Additional observations included an aircraft-mounted photodiode over northeastern Vermont ($44.6^\circ$N, $72.6^\circ$W), radiosonde support from Pittsburg, New Hampshire ($45.05^\circ$N, $71.39^\circ$W), and a ground-based shadow-band video from eastern New Brunswick, Canada ($46.7^\circ$N, $64.8^\circ$W).}
    \label{fig:map}
\end{figure*}

\section{Observation and Data Collection}
\subsection{Observing Sites and Eclipse Conditions}
Observations were conducted during the April 8, 2024 total solar eclipse using ground-based, high-altitude balloon, aircraft, and radiosonde measurements. The primary observing site was Concan, Texas, which lay within the path of totality. At this location, totality lasted from 18:30:13 to 18:34:37 UTC. Two high-altitude balloons carrying photodiode light sensors were launched from Concan, and additional light sensors were deployed on the ground near the launch site.

Supplementary observations were obtained from northeastern Vermont, New Hampshire, and eastern New Brunswick, Canada, as shown in Figure~\ref{fig:map}. An aircraft-mounted photodiode operated at an altitude of approximately 3.6 km over northeastern Vermont. Radiosonde data from a New Hampshire observing team were used to characterize the lower atmosphere near the flight region of the aircraft. In eastern New Brunswick, visual video observations documented shadow bands appearing on snow-covered ground~\cite{seruntine2024}. This video confirms that shadow bands occurred at one location during the eclipse, but the site was geographically separated from both the Texas balloon observations and the Vermont aircraft observations.

\subsection{Instrumentation and Deployment}

Primary light-intensity measurements were obtained using photodiode sensors deployed on high-altitude balloons, at ground level, and on an aircraft. In Concan, two high-altitude balloon payloads were launched, each carrying a 360-degree camera system and two photodiode light sensors. The four balloon-mounted light sensors are designated BL00, BL01, BL02, and BL03. The sensors sampled at approximately 400 Hz and recorded voltage as a function of time, providing high-cadence measurements of changes in solar illumination during the eclipse. At the time of totality, the two balloon payloads were at altitudes of approximately 25 km and 20 km, respectively, placing them well above the PBL. 

The photodiodes used in this study were Hamamatsu sensors with an active area of 12.96~mm$^2$. This collecting area is larger than that used in the 2017 eclipse-ballooning study to reduce instrument noise and provide precise measurements. Ground-based photodiode light sensors were also deployed at the Concan launch site to measure near-surface light-intensity fluctuations. However, as discussed below, cloud cover during totality limited the usefulness of these measurements.

An additional photodiode light sensor was mounted on an aircraft operating over northeastern Vermont at an altitude of approximately 3.6 km. This aircraft measurement provided a geographically separate above-boundary-layer light curve under clearer observing conditions than those available in Concan. Because radiosondes were not launched directly from the aircraft location, radiosonde data from New Hampshire were used as the nearest available atmospheric reference for estimating the PBL height in the aircraft region.

Atmospheric measurements near the primary observing site were obtained using radiosondes launched from Concan. A total of 31 radiosondes were launched over a 30-hour interval beginning approximately 24 hours before totality and continuing for six hours afterwards, with an additional launch during totality. The radiosondes measured temperature, pressure, humidity, wind speed and direction, and altitude, allowing the vertical structure of the lower atmosphere and the approximate height of the PBL to be evaluated during the eclipse. All Concan and New Hampshire radiosonde profiles used in this study were collected as part of the same broader observational effort using the same radiosonde type; the Concan data were recorded at approximately 1~s intervals, while the New Hampshire data were recorded at approximately 2~s intervals. The light-sensor launch information is summarized in Table~\ref{tab:light_sensors}, and the radiosonde launch log is summarized in Table~\ref{tab:radiosondes}.

\begin{table*}[!t]
\centering
{\fontfamily{ptm}\selectfont % Begin Times text block

\caption{\textbf{Data Collection Log from Light Sensors}}
\label{tab:light_sensors}

\begin{tabular}{|l|l|l|l|c|c|l|}
\hline
\textbf{Type} & \textbf{Location} & \makecell{\textbf{Begin Date} \\ \textbf{\& Time (UTC)}} & \makecell{\textbf{End Date} \\ \textbf{\& Time (UTC)}} & \textbf{$^1$Area (mm$^2$)} & \textbf{Altitude (km)} & \textbf{Comments} \\
\hline
Ground GT00 & Concan, TX & 04/08/2024 17:42 & 04/08/2024 18:59 & 12.96 & Ground level & Cloud cover \\
HAB BL00 + BL01 & Concan, TX & 04/08/2024 16:56 & 04/08/2024 17:53 & 12.96 & 25 & Above PBL \\
HAB BL02 + BL03 & Concan, TX & 04/08/2024 16:30 & 04/08/2024 18:18 & 12.96 & 20 & Above PBL \\
Aircraft & NE Vermont & 04/09/2024 19:09 & 04/09/2024 19:40 & 12.96 & 3.6 & Above PBL \\
\hline
\end{tabular}
\vspace{2mm}

\noindent\footnotesize\textit{$^1$The photodiode area for this study was 12.96 mm$^2$, whereas a prior 2017 eclipse-ballooning study reported 1.21~mm$^2$ \cite{madhani2020}, The electronics in the present study were improved to minimize noise.}

\vspace{2mm}
\newcolumntype{Y}[1]{>{\raggedright\arraybackslash}p{#1}}

\caption{\textbf{Data Collection Log from Radiosondes Launched in Concan, Texas}}
\label{tab:radiosondes}

\begin{tabular}{|c|Y{2.8cm}|Y{2.7cm}|Y{2.7cm}|Y{2.3cm}|}
\hline
\textbf{Launch} & \textbf{Launch Time (UTC)} & \textbf{Max Altitude (km)} & \textbf{Surface Temp. (°C)} & \textbf{Complete Profile} \\
\hline

01 & 04/07 18:18 & 32.606 & 25.3 & Y \\
02 & 04/07 19:19 & 32.475 & 26.1 & Y \\
03 & 04/07 20:18 & 32.769 & 27.7 & Y \\
04 & 04/07 21:18 & 33.941 & 28.6 & Y \\
05 & 04/07 22:20 & 33.010 & 29.2 & N \\
06 & 04/07 23:18 & 35.045 & 28.2 & Y \\
07 & 04/08 00:18 & 32.493 & 27.7 & Y \\
08 & 04/08 01:18 & 12.818 & 24.5 & N \\
09 & 04/08 02:18 & 32.210 & 21.0 & Y \\
10 & 04/08 03:18 & 29.745 & 17.3 & Y \\
11 & 04/08 04:18 & 26.185 & 16.1 & Y \\
12 & 04/08 05:18 & 31.283 & 15.4 & Y \\
13 & 04/08 06:18 & 34.363 & 17.0 & Y \\
14 & 04/08 07:18 & 33.154 & 19.9 & Y \\
15 & 04/08 08:18 & 32.967 & 20.6 & Y \\
16 & 04/08 09:18 & 34.180 & 19.3 & Y \\
17 & 04/08 10:18 & 34.602 & 19.3 & Y \\
18 & 04/08 11:18 & 34.244 & 19.6 & Y \\
19 & 04/08 12:18 & 33.583 & 19.7 & Y \\
20 & 04/08 13:18 & 31.486 & 19.8 & Y \\
21 & 04/08 14:18 & 23.343 & 20.4 & Y \\
22 & 04/08 15:18 & 32.733 & 20.7 & Y \\
23 & 04/08 16:18 & 25.725 & 22.8 & N \\
24 & 04/08 17:18 & 33.555 & 24.5 & Y \\
25 & 04/08 18:18 & 29.705 & 23.8 & Y \\
25b & 04/08 18:26 & 30.495 & 23.7 & Y \\
26 & 04/08 19:19 & 34.643 & 23.3 & Y \\
27 & 04/08 20:18 & 33.209 & 23.1 & Y \\
28 & 04/08 21:18 & 30.712 & 21.9 & Y \\
29 & 04/08 22:18 & 32.743 & 24.0 & Y \\
30 & 04/08 23:18 & 32.736 & 25.4 & Y \\
\hline
\end{tabular}

\vspace{1mm}
\begin{minipage}{0.95\textwidth}
\footnotesize\textit{All launches listed here were from Concan, Texas, and used DFM-17 radiosondes. Launch 25b was made during totality. Flights 05 and 08 ended early because of signal loss; Flight 23 was ended early to prepare for the during-totality launch.}
\end{minipage}

} % End fontfamily
\end{table*}

%==========================================

\section{Data Analysis}

\subsection{Light-Curve Processing and Spectrogram Analysis}
\label{sec:curve}
Photodiode voltage time series were analyzed using short-time Fast Fourier Transform (FFT) spectrograms to search for periodic light-intensity fluctuations associated with shadow bands. This approach applies Fourier analysis to consecutive finite-duration segments of the time series, allowing the frequency content of the signal to be examined as a function of time \cite{oppenheim1999,welch1967}. The method is appropriate for this study because shadow band-like signals, if present, are expected to appear as transient or persistent narrow-band features near totality rather than as changes only in the full-duration average power spectrum.

The light sensors sampled at approximately 400 Hz, corresponding to a sampling interval of about 0.0025 s and a Nyquist frequency of approximately 200 Hz. For each time series, the effective sampling frequency was calculated from the median spacing between adjacent time samples rather than assumed to be exactly constant. This accounts for small timing variations in the recorded data before the spectrogram was computed.

Spectrograms were computed by dividing the voltage time series into consecutive 5~s segments and applying an FFT to each segment. At the measured sampling rate, a 5~s segment contains approximately 2000 samples. The resulting frequency-bin spacing is therefore 
\begin{equation}
\Delta f = \frac{f_s}{N}
\approx \frac{400~\mathrm{Hz}}{2000}
\approx 0.20~\mathrm{Hz},
\label{eq}
\end{equation}
where $f_s$ is the sampling frequency and $N$ is the number of samples in each FFT segment. This frequency resolution is sufficient to resolve narrow band features near the 4.5~Hz signal reported in the 2017 eclipse-ballooning study \cite{madhani2020}, while retaining enough time resolution to evaluate whether such features appear before, during or after totality.

Adjacent FFT segments were overlapped by 50\%, corresponding to a 2.5 s step between neighboring spectrogram columns. This overlap reduces the chance that a short-lived feature would fall between adjacent time bins and provides a smoother time-frequency representation while preserving the time resolution. The same segment length, overlap, windowing, and display limits were used for the high-altitude balloon, aircraft, and ground-based photodiode data so that the results from different observing platforms could be compared directly. Although the FFT was computed over the full available frequency range up to the Nyquist frequency, the main figures show 0--25 Hz because the frequencies of interest for comparison with previous shadow band measurements lie well within this interval.

Before spectrogram generation, the high-altitude balloon light curves were cleaned to remove nonphysical voltage drops caused by intermittent obstruction of the photodiode by the payload line. A bidirectional exponentially weighted moving average was used to estimate the slowly varying eclipse light curve. Sharp deviations from this trend were removed and replaced using linear interpolation, followed by a second smoothing and interpolation pass. This procedure was designed to suppress short-duration obstruction artifacts while preserving the gradual eclipse profile and any small-amplitude periodic structure.

The effect of this processing is shown for the BL00 photodiode in Figure~\ref{fig:BL00}. The same spectrogram procedure was also applied to synthetic signal-injection tests, described in Section~\ref{sec:inj}, to evaluate whether shadow band-like periodic signals would remain detectable after the adopted processing.

\subsection{Signal-Injection and Detection-Limit Test}
\label{sec:inj}
To evaluate how a shadow band-like periodic signal would appear in the adopted spectrogram analysis, synthetic sinusoidal signals were injected into the BL00 voltage time series before applying the cleaning procedure. The injected data were then processed through the same cleaning and spectrogram pipeline used for the observational light curves. This test was designed to determine whether the processing and plotting procedure described in Section~\ref{sec:curve} would preserve a narrow-band signal comparable to that reported in the 2017 eclipse-ballooning study. The injected signal frequency was set to 4.5 Hz, matching the persistent frequency reported in that earlier work \cite{madhani2020}.

The injected light curve was defined as
\begin{equation}
V_{\mathrm{inj}}(t)
=
V_{\mathrm{raw}}(t)
+
A\,\sin(2\pi f_{0}t),
\label{eq:signal_injection}
\end{equation}
where $V_{\mathrm{raw}}(t)$ is the raw BL00 voltage time series,
$A$ is the injected signal amplitude, and $f_{0}=4.5~\mathrm{Hz}$. The signal-injected time series was then passed through the same cleaning and spectrogram procedure used for the observational data. This ordering was chosen to test whether a real periodic signal already present in the raw photodiode data would survive the artifact-removal procedure. The sinusoidal signal was added only before and after totality, and was excluded during totality itself. This choice was made to mimic the timing of the previously reported 4.5 Hz signal, which appeared outside totality rather than during the entire eclipse process. The amplitude scale for the injection test was chosen as an order-of-magnitude reference rather than as a measured 2024 shadow band-amplitude. Previous photoelectric observations found shadow band fluctuations at the level of only a few percent of the background illumination, with Marschall et al. reporting weak 2-3\% fluctuations during the 1980 total eclipse \cite{marschall1984}. In the BL00 light curve, the photodiode voltage outside totality is of order a few volts, so a 1\% intensity modulation corresponds to a voltage fluctuation of order $10^{-2}$ V. We therefore adopted $0.03$ V as a representative reference fluctuation amplitude. Synthetic signals were injected at $A=0.0030$ V, $0.0009$ V, and $0.0003$ V, corresponding to 10\%, 3\%, and 1\% of this reference amplitude. These values test whether the analysis could recover signals substantially smaller than a percent-level shadow band modulation.

Figure~\ref{fig:BL00inj} shows the resulting cleaned spectrograms for the three injected amplitudes. The vertical dashed lines mark the totality interval. The signal is most visible in the 10\% case and remains identifiable in the 3\% case. At the 1\% level, the feature is nearly undetectable visibly, but it still produces enhanced power near 4.5 Hz relative to the no-injection baseline. This behavior is shown in the mean power spectrum computed over a 15 s interval before totality. The signals produce a localized peak at 4.5 Hz, while the baseline spectrum does not show a comparable feature. The signal-injection test therefore demonstrates that a persistent narrow-band signal at or above the tested amplitudes would produce a recognizable signature in the adopted spectrogram and power-spectrum analysis.

No comparable transient or persistent signal feature is present in the actual BL00 spectrogram or in the other 2024 high-altitude balloon and aircraft light-sensor data. The absence of such a feature in the observational data is therefore not simply a consequence of the spectrogram display or the cleaning procedure. Instead, within the sensitivity explored by this injection test, the 2024 data show no detectable shadow band-like periodic signal above the PBL.

\subsection{Photodiode Data Spectrogram Results}
\label{sec:discuss}

\begin{figure}
    \centering

    \includegraphics[width=\linewidth]{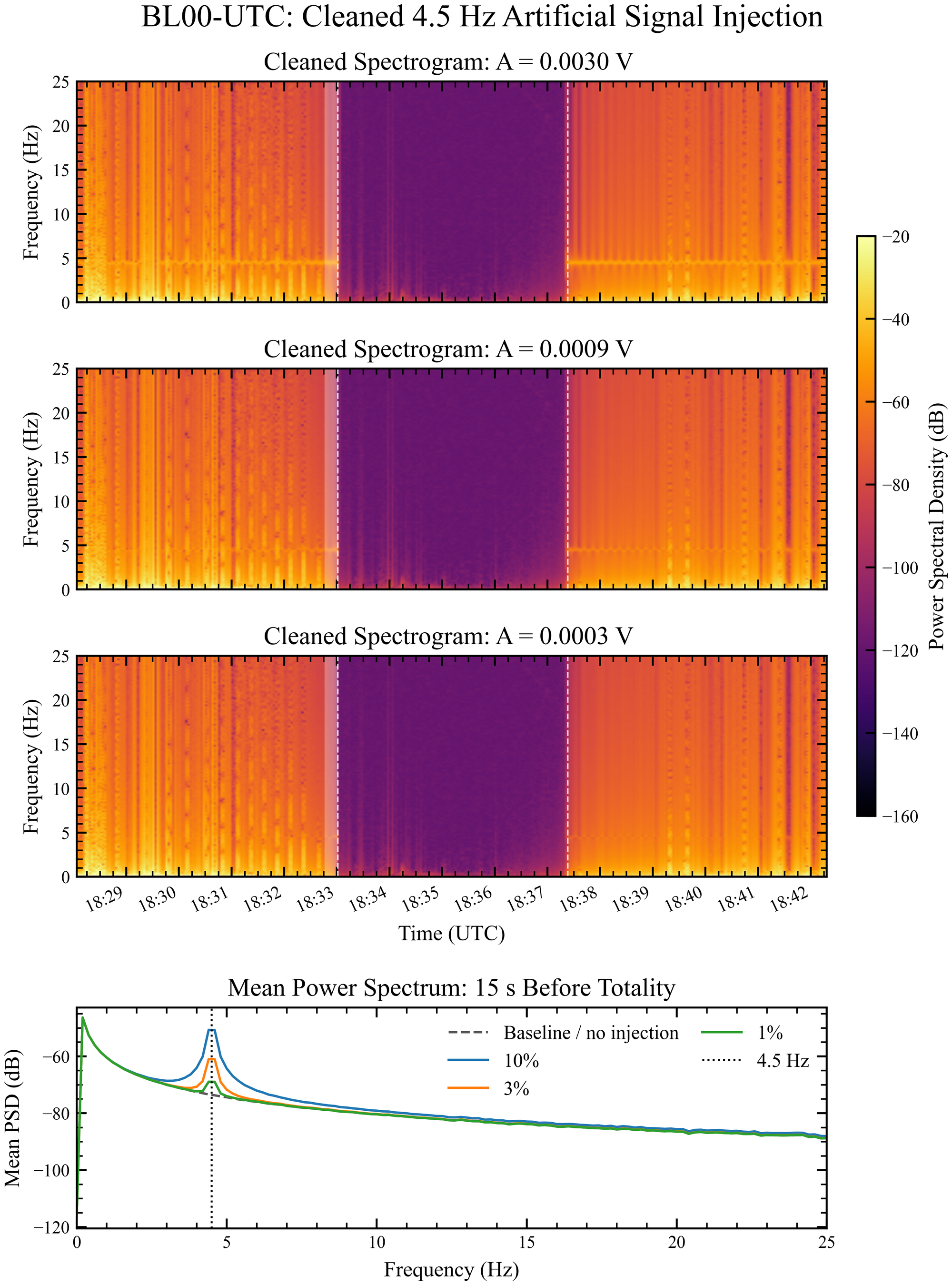}
    \caption{\textbf{Signal-injection test for the BL00 photodiode data.}
Synthetic 4.5 Hz sinusoidal signals were added to the BL00 voltage time series before and after totality, but not during totality, and the signal-injected data were then processed through the same cleaning and spectrogram pipeline used for the observational data. The three spectrogram panels show injected amplitudes of $A=0.0030$ V, $0.0009$ V, and $0.0003$ V. Vertical dashed lines mark the totality interval. The lower panel shows the mean power spectrum computed over a 15 s interval before totality (highlighted in the spectrogram), with the dotted vertical line marking 4.5 Hz. The injected signals produce localized power near 4.5 Hz, demonstrating how a persistent narrow-band signal would appear in the adopted analysis.}
    \label{fig:BL00inj}
\end{figure}

After establishing the spectrogram procedure and signal-injection test, the same analysis was applied to the 2024 photodiode measurements from the high-altitude balloons, aircraft, and ground-based sensors. The purpose of this section is to compare the observed spectrograms with the injected-signal behavior described in Section~\ref{sec:inj}. A shadow band-like periodic signal would be expected to appear as a localized enhancement in power at a fixed frequency, similar to what is shown in Figure~\ref{fig:BL00inj}.

\begin{figure}

\centering

\begin{minipage}[t]{\columnwidth}
\centering
\includegraphics[width=0.95\linewidth]{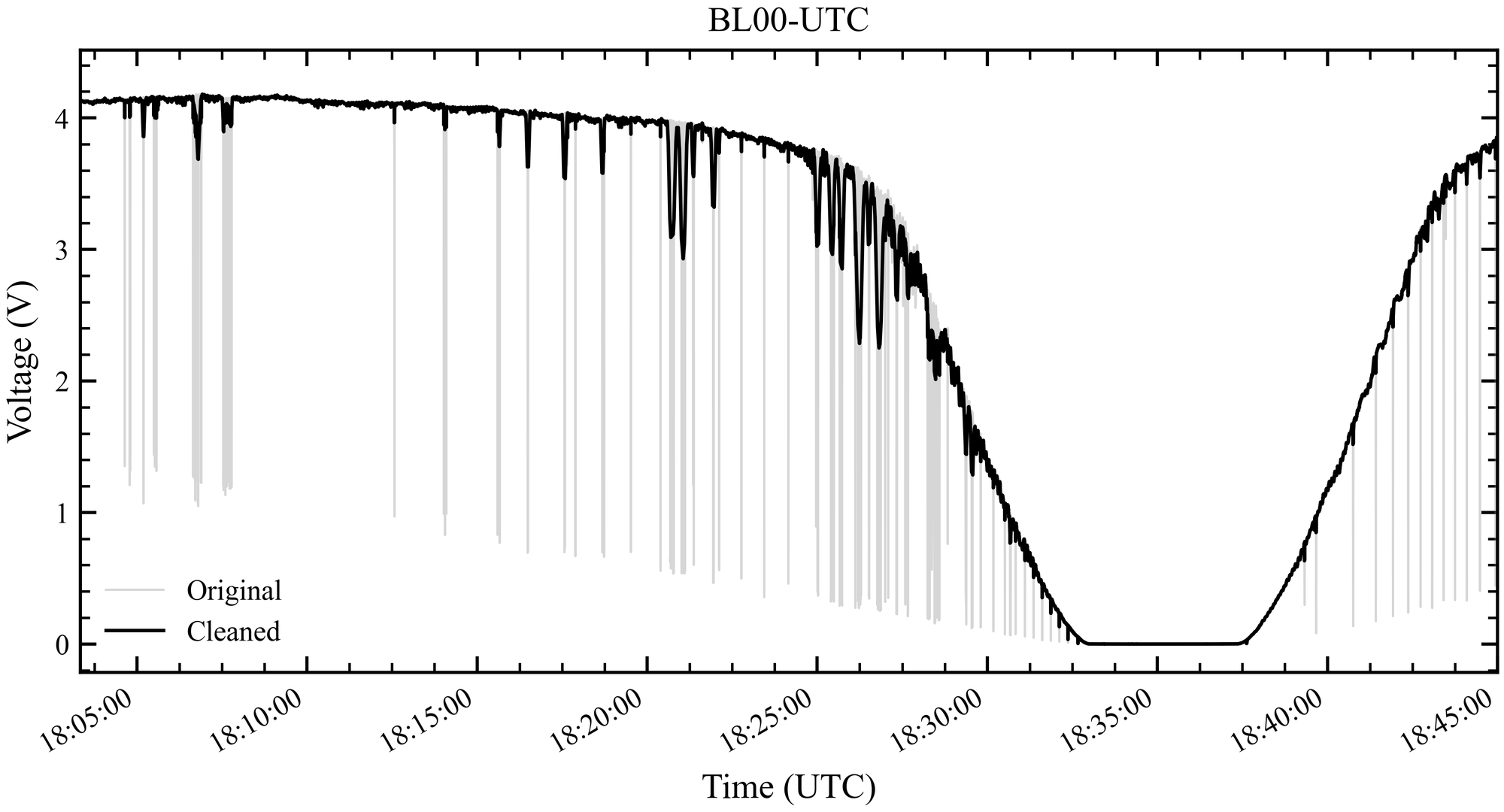}
\par\small \textbf{(a) Light curve BL00}
\end{minipage}

\vspace{6pt}

\begin{minipage}[t]{\columnwidth}
\centering
\includegraphics[width=0.95\linewidth]{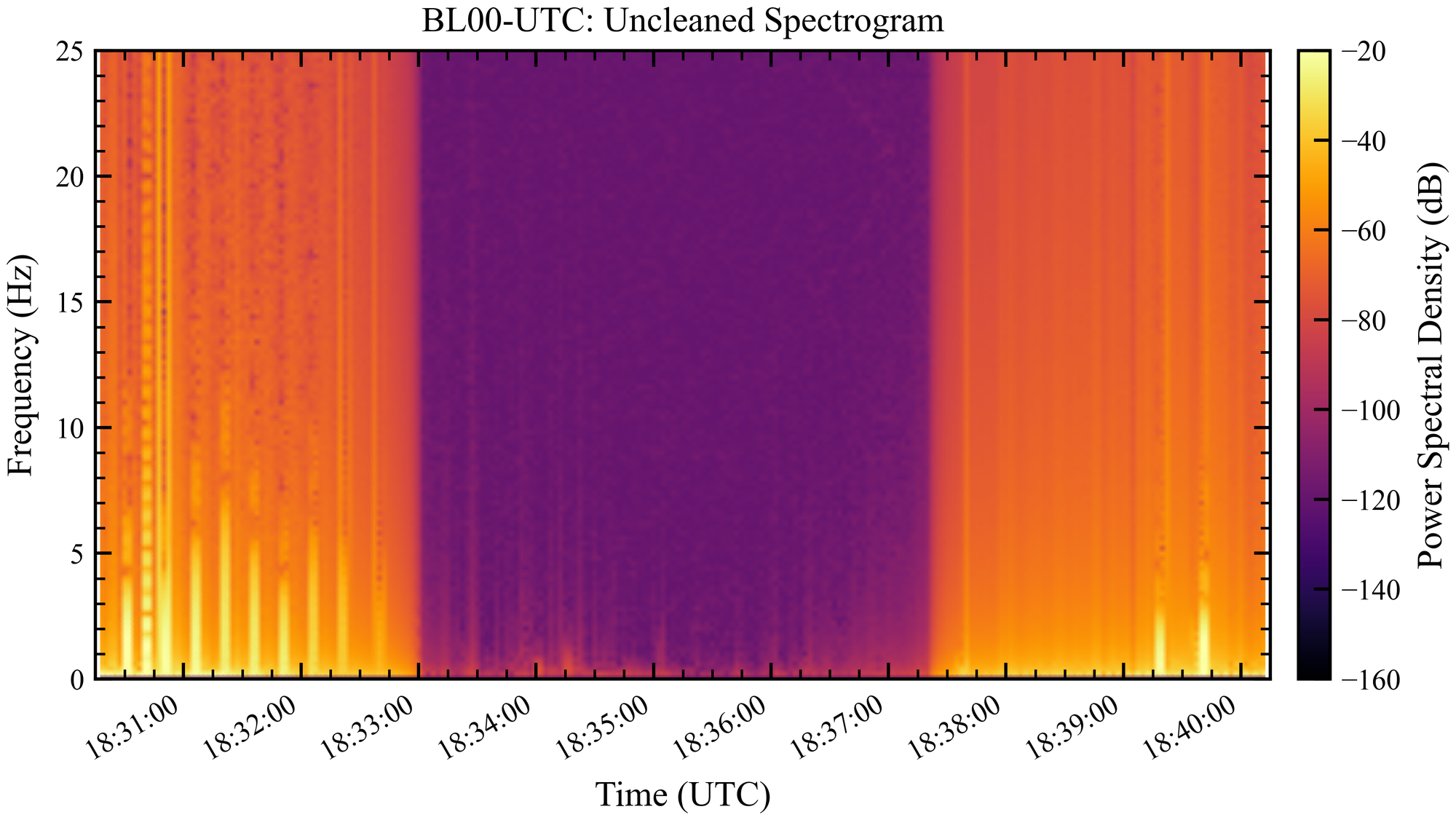}
\par\small \textbf{(b) Raw spectrogram BL00}
\end{minipage}

\vspace{6pt}

\begin{minipage}[t]{\columnwidth}
\centering
\includegraphics[width=0.95\linewidth]{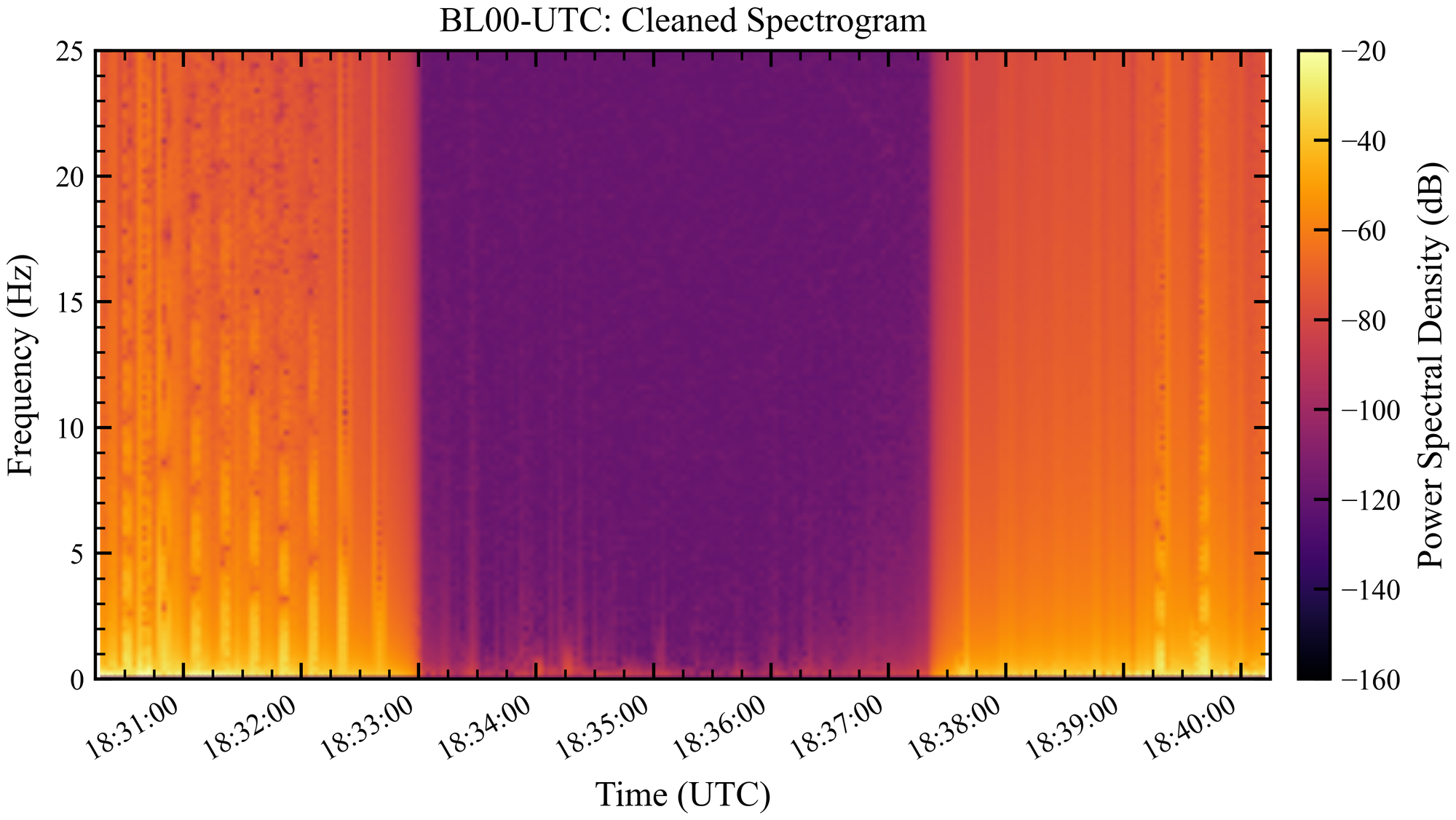}
\par\small \textbf{(c) Cleaned spectrogram BL00}
\end{minipage}

\caption{\textbf{Light-curve cleaning and spectrogram construction for the HAB BL00 photodiode.}
(a) Raw and cleaned voltage time series. The sharp downward excursions in the raw signal are caused by intermittent payload-line obstruction of the photodiode. (b) Spectrogram of the uncleaned BL00 light curve, showing these artifacts. (c) Spectrogram after artifact suppression. The cleaned spectrogram shows no persistent narrow-band frequency feature comparable to a shadow-band-like signal.}
\label{fig:BL00}
\end{figure}

Figure~\ref{fig:BL00} shows the raw and cleaned BL00 light curve, together with the corresponding raw and cleaned spectrograms. The uncleaned data contain sharp voltage dips associated with intermittent payload-line obstruction of the photodiode. These artifacts introduce nonphysical structure into the raw spectrogram. After cleaning, the eclipse light-curve morphology is preserved while the strongest artifact-related features are suppressed but not eliminated. The cleaned BL00 spectrogram does not show a persistent narrow-band signal comparable to the injected signal shown in Figure~\ref{fig:BL00inj}. In particular, no continuous or semi-continuous horizontal feature is present throughout the entire frequency range inspected before or after totality. The other HAB photodiodes, BL01--BL03, show similar behavior, with no persistent frequency signature detected in the cleaned spectrograms. These additional HAB spectrograms are provided in Appendix Figures~\ref{fig:BL01}--\ref{fig:BL03}.

The aircraft-mounted photodiode data were analyzed using the same spectrogram procedure. The aircraft measurement provides an independent above-boundary-layer light curve from northeastern Vermont, geographically separated from the Concan HAB observations. As in the HAB data, the aircraft spectrogram does not show a persistent narrow-band signal near 4.5 Hz or any other stable frequency feature that resembles the injected-signal response. This result indicates that no detectable shadow band-like periodic signal was present in the aircraft data during the sampled interval.

Cloud cover during totality strongly affected the Concan ground-based photodiode light curve, reducing its usefulness for detecting near-surface shadow band signatures. As a result, the data cannot be used to determine whether shadow bands were present or absent near the surface at Concan. Because the Concan ground-based photodiode data were affected by cloud cover, the radiosonde measurements at that site could not be paired with reliable ground-level light-intensity measurements to test directly for a relationship between boundary-layer structure and near-surface shadow band behavior. The photodiode non-detection result therefore applies to the high-altitude balloon and aircraft observations, while the near-surface behavior at Concan remains unconstrained by the photodiode data.

\begin{figure}
\centering

\includegraphics[width=0.92\columnwidth]{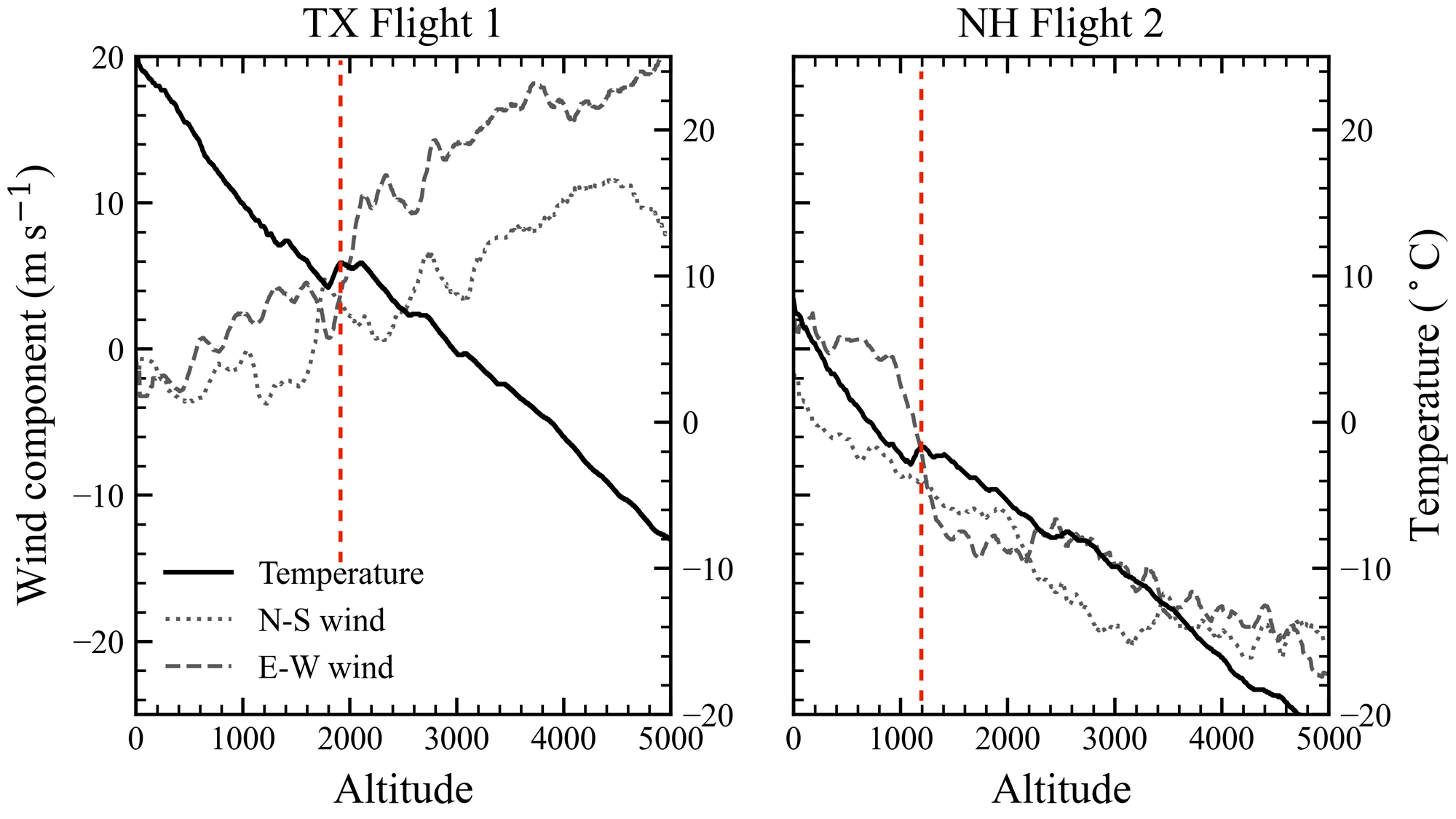}
\par\smallskip
{\small (a) Pre-eclipse conditions}

\vspace{0.8em}

\includegraphics[width=0.92\columnwidth]{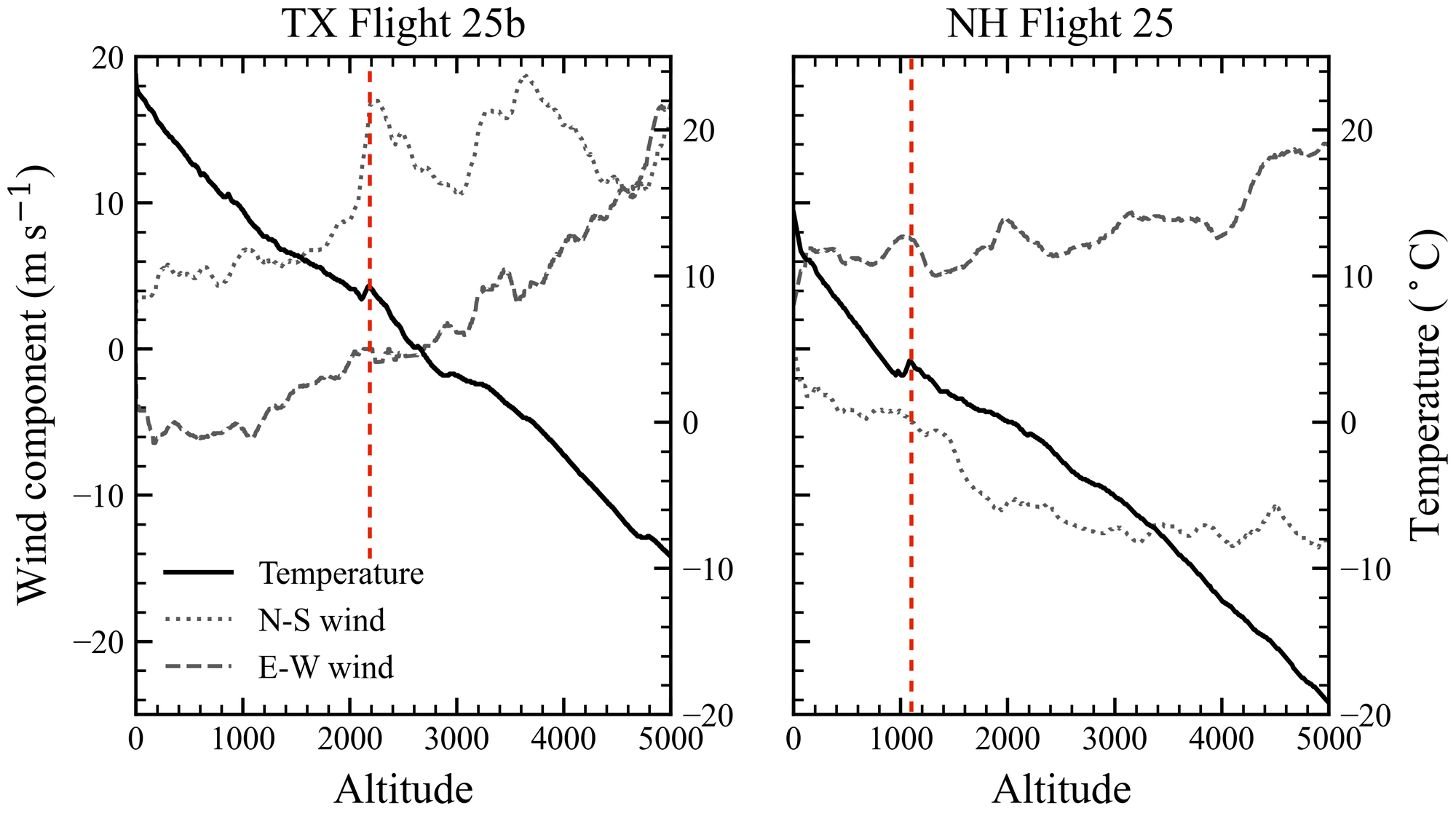}
\par\smallskip
{\small (b) Closest-to-eclipse conditions}

\vspace{0.8em}

\includegraphics[width=0.92\columnwidth]{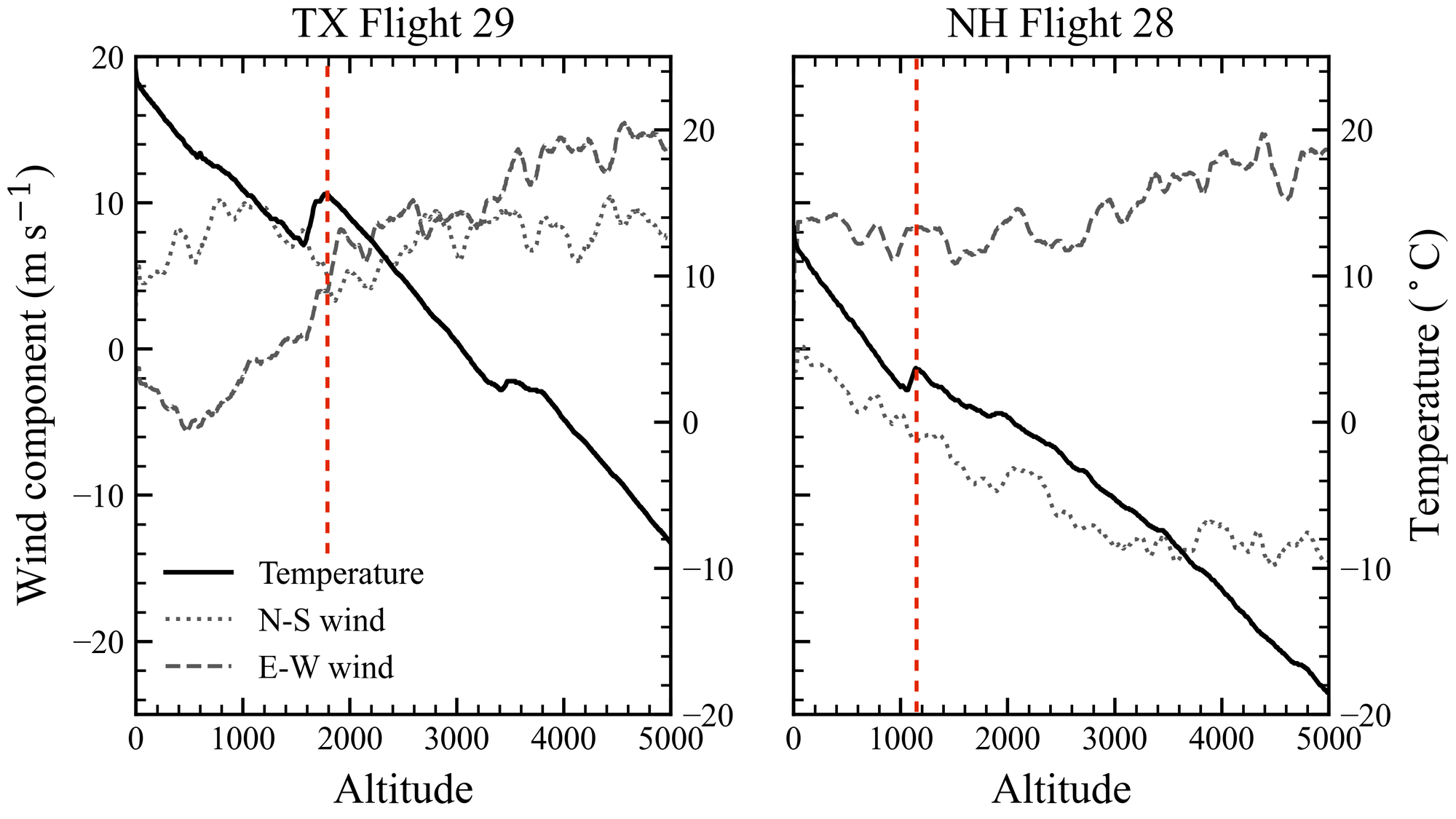}
\par\smallskip
{\small (c) Post-eclipse conditions}

\caption{\textbf{Representative radiosonde temperature and wind profiles used to estimate the PBL height.}
Profiles are shown for paired Concan, Texas (left) and Pittsburg, New Hampshire (right) radiosonde flights before the eclipse, closest to the eclipse, and after the eclipse. Temperature inversions were used to identify stable layers capping the near-surface mixed layer. The inferred PBL heights remained well below the HAB altitudes over Concan and below the aircraft altitude over northeastern Vermont.}
\label{fig:radiosonde_profiles_combined}

\end{figure}

\subsection{Radiosonde Analysis and Planetary Boundary Layer Determination}
Radiosonde data were analyzed to determine whether the photodiode measurements were obtained within or above the lower turbulent atmosphere. This distinction is central to the interpretation of the light-sensor results because one proposed explanation for shadow bands attributes them to refractive distortions produced by turbulence within the PBL. The PBL is the portion of the lower atmosphere directly influenced by surface heating, friction, and turbulent mixing, and its height can vary substantially with time of day, surface conditions, and atmospheric stability \cite{stull1988}.

The PBL height was estimated from radiosonde temperature profiles by identifying the lowest pronounced stable layer or temperature inversion that is at the top of the near-surface mixed layer. In a daytime convective boundary layer, the well-mixed region near the surface is often capped by a stable layer in which temperature or potential temperature increases with height. This inversion inhibits vertical mixing and provides an appropriate marker for the transition from the turbulent boundary layer to the more weakly mixed free atmosphere \cite{stull1988, seibert2000}. Wind speed and wind direction profiles were also examined because changes in wind shear or directional structure can provide supporting evidence for the transition between the turbulent mixed layer and the more stable air above the PBL \cite{stull1988, seibert2000}. Because radiosonde-based PBL estimates rely on features in the temperature profile rather than a direct measurement of the boundary layer, the inferred PBL heights are subject to uncertainty and should be regarded as approximate rather than sharply defined boundaries \cite{seidel2010}. Based on the altitude spacing between consecutive samples, the Concan, Texas radiosonde data had an effective lower-atmosphere vertical resolution of approximately 6~m, while the Pittsburg, New Hampshire radiosonde data had an effective lower-atmosphere vertical resolution of approximately 7~m.

For the Concan observing site, radiosonde profiles were examined before, during, and after totality. The temperature profiles show a near-surface temperature inversion well below the HAB payloads. Around totality, the inferred PBL height remained below approximately 2.5 km. The HAB photodiode payloads were located at altitudes of approximately 20 km and 25 km at the time of totality, far above the inferred boundary-layer height. This large altitude separation indicates that the HAB light curves were not obtained within the lower turbulent layer sampled by the radiosondes.

The aircraft-mounted photodiode measurement was interpreted using radiosonde data from New Hampshire, the nearest available atmospheric reference for the northeastern Vermont aircraft flight region. These profiles indicate that the PBL height in the regional lower atmosphere remained below approximately 1.5 km near the eclipse period. The aircraft operated at an altitude of approximately 3.6 km, placing it above the inferred boundary layer. Because the radiosonde launch site was not co-located with the aircraft flight path, this estimate carries greater spatial uncertainty than the Concan HAB comparison. Additionally, the New Hampshire radiosonde profiles were not co-located with ground-based light sensors. However, the available profiles still indicate that the aircraft measurement was likely obtained above the lower turbulent atmosphere.

Figure~\ref{fig:radiosonde_profiles_combined} shows the representative radiosonde profiles used to estimate the PBL height for the Concan and New Hampshire observing regions. Temperature is plotted with wind speed and wind direction to show whether thermodynamic changes near the inferred PBL height are accompanied by changes in the vertical wind structure. The inferred PBL heights are well below the HAB altitudes and below the aircraft altitude. These atmospheric measurements therefore support the interpretation that the high-altitude balloon and aircraft photodiode non-detections correspond to observations made above the PBL. The radiosonde analysis does not determine whether shadow bands were present near the surface at Concan, especially because cloud cover limited the ground-based photodiode data. Instead, it establishes the atmospheric context needed to interpret the above-boundary-layer light-sensor measurements.

%========================================

\section{Discussion and Summary}
\subsection{Interpretation of the Above-Boundary-Layer Non-Detections}
The primary result of this study is the absence of a detectable shadow band-like periodic signal in the 2024 above-boundary-layer photodiode measurements. The high-altitude balloon sensors were located at approximately 20 km and 25 km during totality, while the aircraft-mounted sensor operated at approximately 3.6 km. Radiosonde profiles indicate that these platforms were above the inferred planetary boundary layer during the relevant observing intervals. Therefore, the HAB and aircraft measurements provide tests of whether a persistent, detectable light-intensity fluctuation was present above the lower turbulent atmosphere.

The spectrogram analysis did not reveal a persistent 4.5 Hz signal, or any other stable narrow-band feature resembling the injected-signal response, in the HAB or aircraft photodiode data. This result is significant because the signal-injection test showed that a persistent 4.5 Hz signal at or above the tested amplitudes would produce a recognizable feature in the adopted spectrogram and power-spectrum analysis. The absence of such a feature in the observational data is therefore not simply a consequence of the cleaning procedure, the chosen spectrogram display, or the frequency range shown in the figures.

These results do not rule out the occurrence of shadow bands at all locations during the April 8, 2024 eclipse. Instead, they constrain the presence of detectable shadow band-like periodic signals above the PBL at the locations and times sampled by the HAB and aircraft measurements. The cloud-affected Concan ground-based photodiode data do not provide a reliable constraint on near-surface shadow band behavior, so the present analysis cannot determine whether shadow bands were present on the ground at Concan.

The above-boundary-layer non-detections are most relevant to interpretations in which shadow band-like signals should persist above the lower turbulent atmosphere. If shadow bands were produced by a mechanism that generated a repeatable photodiode-detectable signal above the PBL under the sampled observing conditions, then a corresponding narrow-band feature would be expected in the HAB or aircraft spectrograms. No such feature was observed. The 2024 data therefore do not provide support for a persistent above-PBL photodiode signature comparable to that reported in the 2017 eclipse-ballooning study.

\subsection{Comparison with Previous Eclipse Observations}
The 2024 above-boundary-layer non-detections differ from the result reported during the August 21, 2017 total solar eclipse. In that study, a persistent 4.5 Hz signal was reported in both ground-based and high-altitude balloon photodiode measurements, with the high-altitude balloon payload located at approximately 25 km. Because this signal was observed above the lower turbulent atmosphere, it was interpreted as possible evidence for a shadow band-like optical signature above the PBL \cite{madhani2020}. In the present study, no comparable 4.5 Hz signal was detected in the HAB photodiode data from Concan or in the aircraft photodiode data from northeastern Vermont.

The difference between the 2017 and 2024 results may reflect the intermittent and site-dependent nature of shadow bands. Historical observations show that shadow bands are not uniformly reported at every eclipse or at every observing location, and their timing, visibility, spacing, and motion vary substantially between reports \cite{ranyard1879, rotch1908}. The absence of a detectable signal in the 2024 above-PBL measurements therefore does not require that shadow bands were absent everywhere along the eclipse path. Instead, it shows that a persistent photodiode-detectable signal comparable to the 2017 4.5 Hz feature was not present in the above-boundary-layer measurements collected in this study.

This distinction is consistent with the eastern New Brunswick video observation. Visual video evidence from eastern New Brunswick confirms that shadow bands were visible at one ground location during the April 8, 2024 eclipse. An FFT analysis of the New Brunswick video showed a dominant temporal frequency of approximately $2.7 \pm 0.26~\mathrm{Hz}$. This frequency is lower than the 4.5~Hz signal reported in 2017 and was obtained from a different type of observation: a ground-based video recording of visible shadow band motion on snow, rather than a photodiode voltage time series.

Because the New Brunswick observation was geographically separated from the Concan and Vermont photodiode measurements, it cannot be used to infer whether shadow bands were present near the primary observing sites. The New Brunswick site was approximately 3500 km from Concan and approximately 540 km from the Vermont aircraft flight region. Local atmospheric conditions, surface properties, viewing geometry, and measurement method all differed from those of the photodiode observations. The video FFT result therefore provides useful evidence that shadow bands occurred somewhere along the April 8 eclipse path, but it is not directly comparable to the HAB and aircraft spectrogram non-detections.

Taken together, the comparison with previous and ancillary observations suggests that the 2024 results should be interpreted as an above-boundary-layer non-detection under the sampled observing conditions, rather than as a claim that shadow bands were absent throughout the eclipse path.

\subsection{Limitations and Implications for Future Observations}
Several observational limitations affect the interpretation of the 2024 non-detections. Cloud cover at Concan prevented the ground-based photodiode data from providing a reliable constraint on near-surface shadow band activity at the primary observing site. This also prevented a direct comparison between local radiosonde-derived boundary-layer structure, near-surface light-intensity fluctuations, and the HAB measurements above the PBL. 

The aircraft measurement provides an independent above-boundary-layer photodiode data set, but it was not paired with co-located ground-based photodiode measurements or radiosonde launches directly beneath the flight path. The New Hampshire radiosonde profiles provide the nearest available atmospheric context for the Vermont aircraft region, but they should be interpreted as regional rather than co-located measurements.

Future eclipse observations should prioritize co-located measurements above and below the PBL. An ideal observing strategy would combine ground-based photodiodes, calibrated video, above-boundary-layer photodiode measurements from aircraft or HABs, and radiosonde launches near totality at the same observing sites. In addition, these strategies should be implemented at multiple locations along the eclipse path. This configuration would allow future studies to determine whether shadow band detections are linked to local boundary-layer structure, above-PBL optical effects, or both.

\subsection{Summary}

The April 8, 2024 observations provide a multi-platform test for shadow band-like light-intensity fluctuations above the lower turbulent atmosphere. The main findings are as follows:

\begin{itemize}
\item No persistent 4.5 Hz signal, or other stable narrow-band feature resembling the injected test signals, was detected in the HAB or aircraft photodiode measurements.
\item Signal-injection tests show that the adopted cleaning and spectrogram procedure would recover persistent 4.5 Hz signals at the tested amplitudes.
\item Radiosonde profiles indicate that the HAB and aircraft photodiode measurements were obtained above the inferred PBL, so the non-detections apply most directly to above-boundary-layer observing conditions.
\item Cloud cover at Concan prevented the ground-based photodiode data from constraining near-surface shadow band activity at the primary observing site.
\item The New Brunswick video observation confirms that shadow bands occurred at one location along the April 8 eclipse path, but its geographic separation from Concan and Vermont prevents direct comparison with the photodiode non-detections.
\end{itemize}

Taken together, the 2024 results do not support the presence of a persistent above-PBL photodiode signature comparable to the 4.5 Hz signal reported during the 2017 eclipse-ballooning study. However, because the near-surface measurements at Concan were limited by cloud cover and the New Brunswick visual detection occurred far from the photodiode observing sites, the results should be interpreted as an above-boundary-layer non-detection under the sampled observing conditions rather than as a path-wide absence of shadow bands.

\section*{Acknowledgments}
We thank Gary and Margaret Miller for flying their plane over northeast Vermont while one of us (JBP) pointed our PL00 light sensor at the eclipsed Sun. We thank Cliff Seruntine, who took the extraordinary video of shadow bands in East Brunswick, Canada, and posted it on youtube. 
We thank and acknowledge Eric Kelsey and Genevieve Picciano, who made the New Hampshire radiosonde data available. And we thank Jerry Baker for his assistance with recovering a test flight payload. We gratefully acknowledge financial support from the NASA Nationwide Eclipse Ballooning Program, the NASA Space Grant Foundation, the NASA Pennsylvania Space Grant Consortium, and the University of Pittsburgh’s Dietrich School, Frederick Honors College, Department of Physics and Astronomy, and Allegheny Observatory.

\appendix

\renewcommand{\thefigure}{A\arabic{figure}}
\renewcommand{\theHfigure}{A\arabic{figure}}
\setcounter{figure}{0}
\section{Additional Light Curve and Spectrogram Results from other Light Sensors}

\noindent
For completeness, this appendix presents additional light curves and spectrograms for the remaining observational platforms. The figures first show the remaining HAB sensors, followed by the aircraft and ground-based sensors. All spectrograms were generated using the same procedure described in Section~\ref{sec:curve}. The cleaning algorithm was applied only to the HAB data; the aircraft and ground data are shown without cleaning.

%\clearpage

\begin{figure}

\centering

\begin{minipage}[t]{\columnwidth}
\centering
\includegraphics[width=0.95\linewidth]{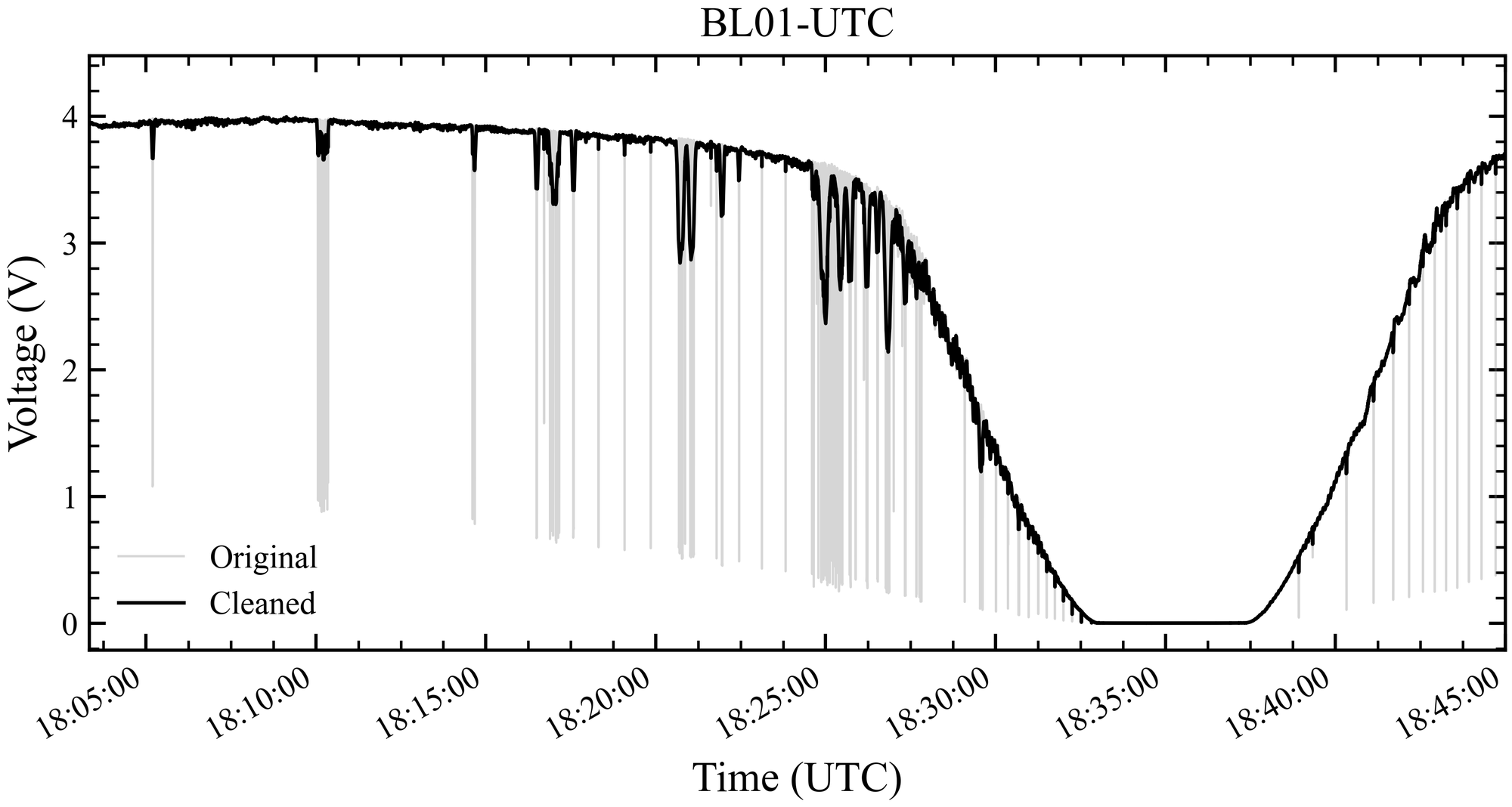}
\par\small \textbf{(a) Light curve BL01}
\end{minipage}

\vspace{6pt}

\begin{minipage}[t]{\columnwidth}
\centering
\includegraphics[width=0.95\linewidth]{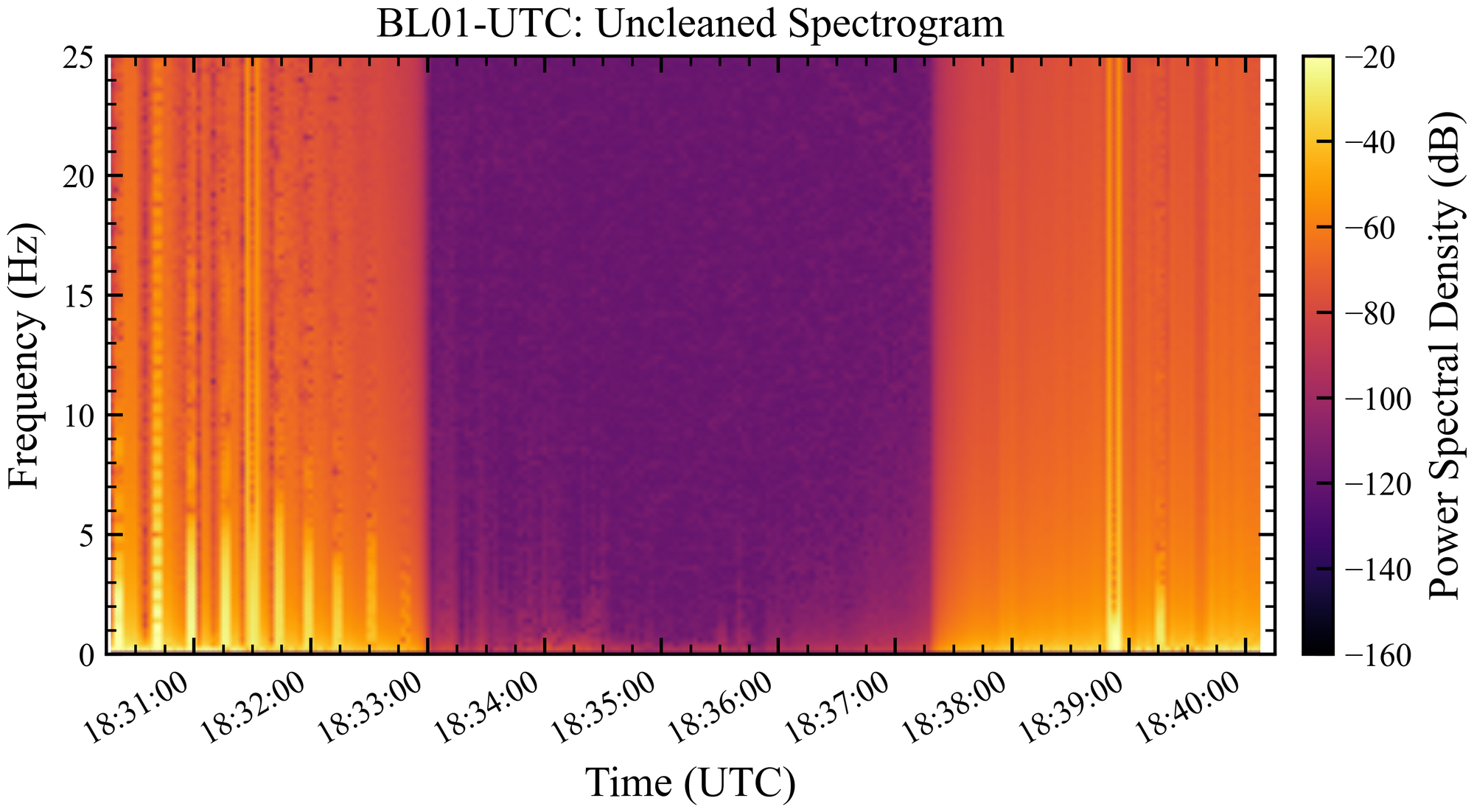}
\par\small \textbf{(b) Raw spectrogram BL01}
\end{minipage}

\vspace{6pt}

\begin{minipage}[t]{\columnwidth}
\centering
\includegraphics[width=0.95\linewidth]{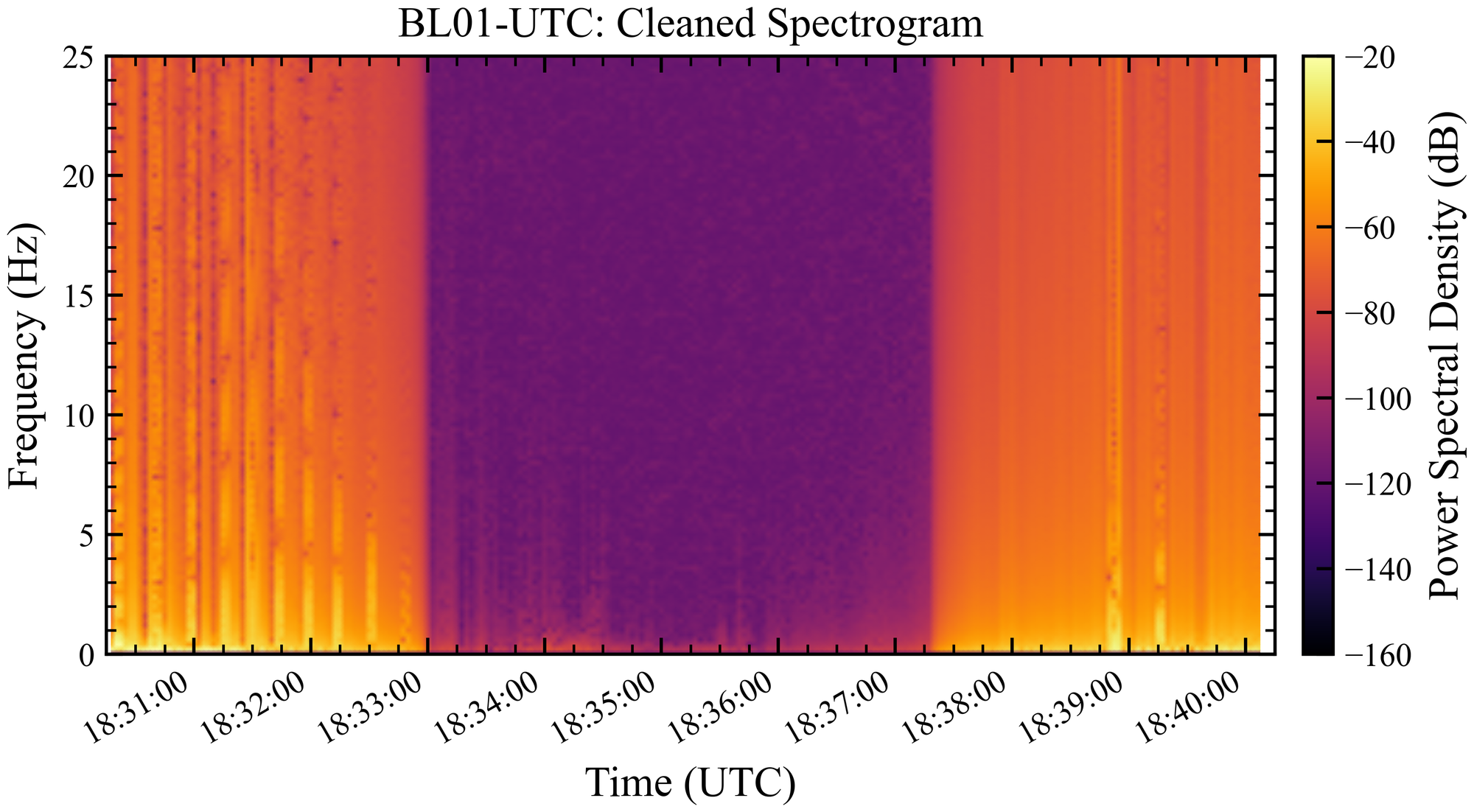}
\par\small \textbf{(c) Cleaned spectrogram BL01}
\end{minipage}

\caption{\textbf{Light-curve cleaning and spectrogram construction for the HAB BL01 photodiode.}
(a) Raw and cleaned voltage time series. (b) Spectrogram before artifact suppression. (c) Spectrogram after artifact suppression. The same cleaning and spectrogram procedure used for BL00 was applied to BL01. No persistent narrow-band frequency feature comparable to a shadow band-like signal is present in the cleaned spectrogram.}
\label{fig:BL01}

\end{figure}

%-----

\begin{figure}

\centering

\begin{minipage}[t]{\columnwidth}
\centering
\includegraphics[width=0.95\linewidth]{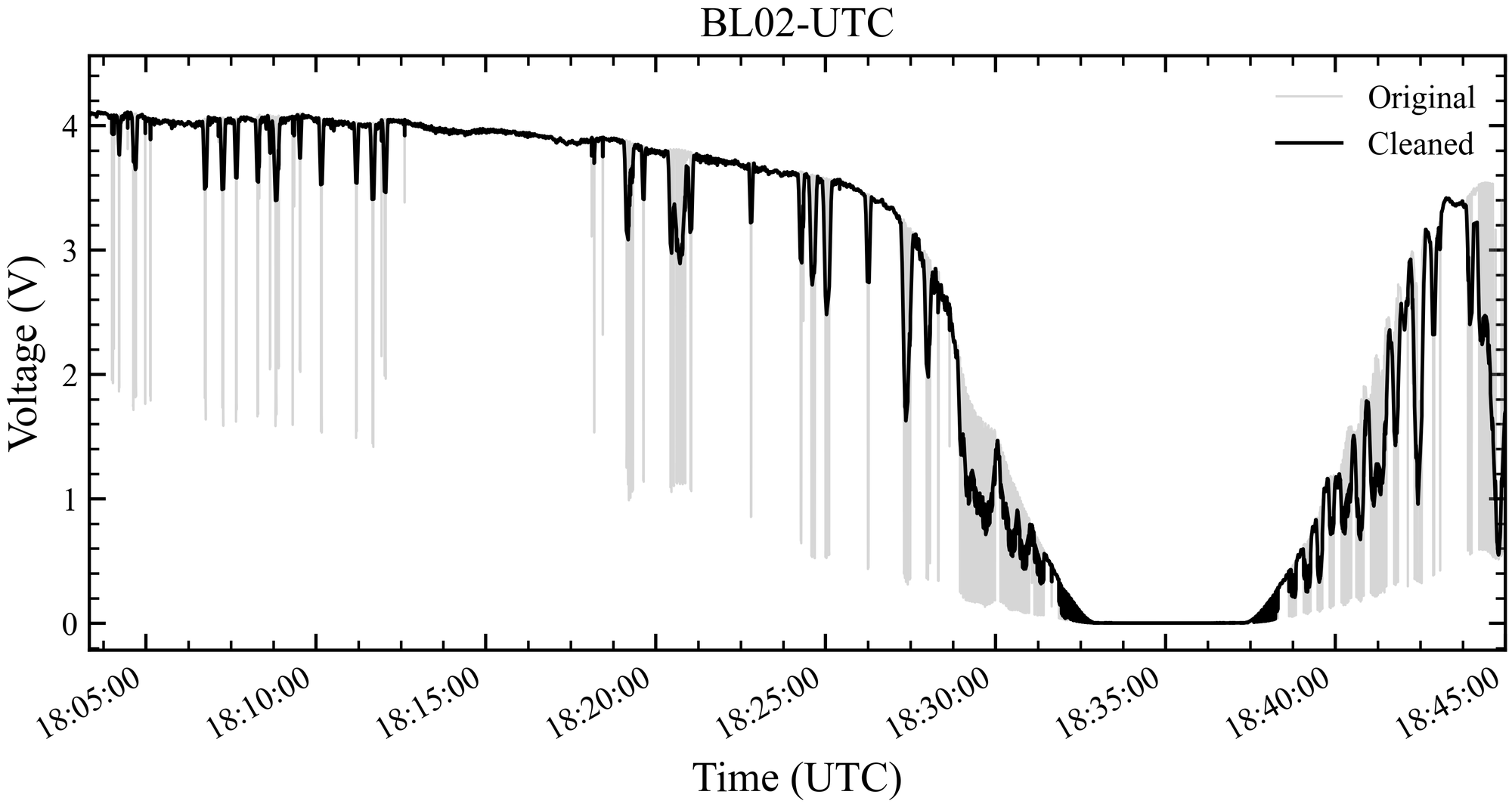}
\par\small \textbf{(a) Light curve BL02}
\end{minipage}

\vspace{6pt}

\begin{minipage}[t]{\columnwidth}
\centering
\includegraphics[width=0.95\linewidth]{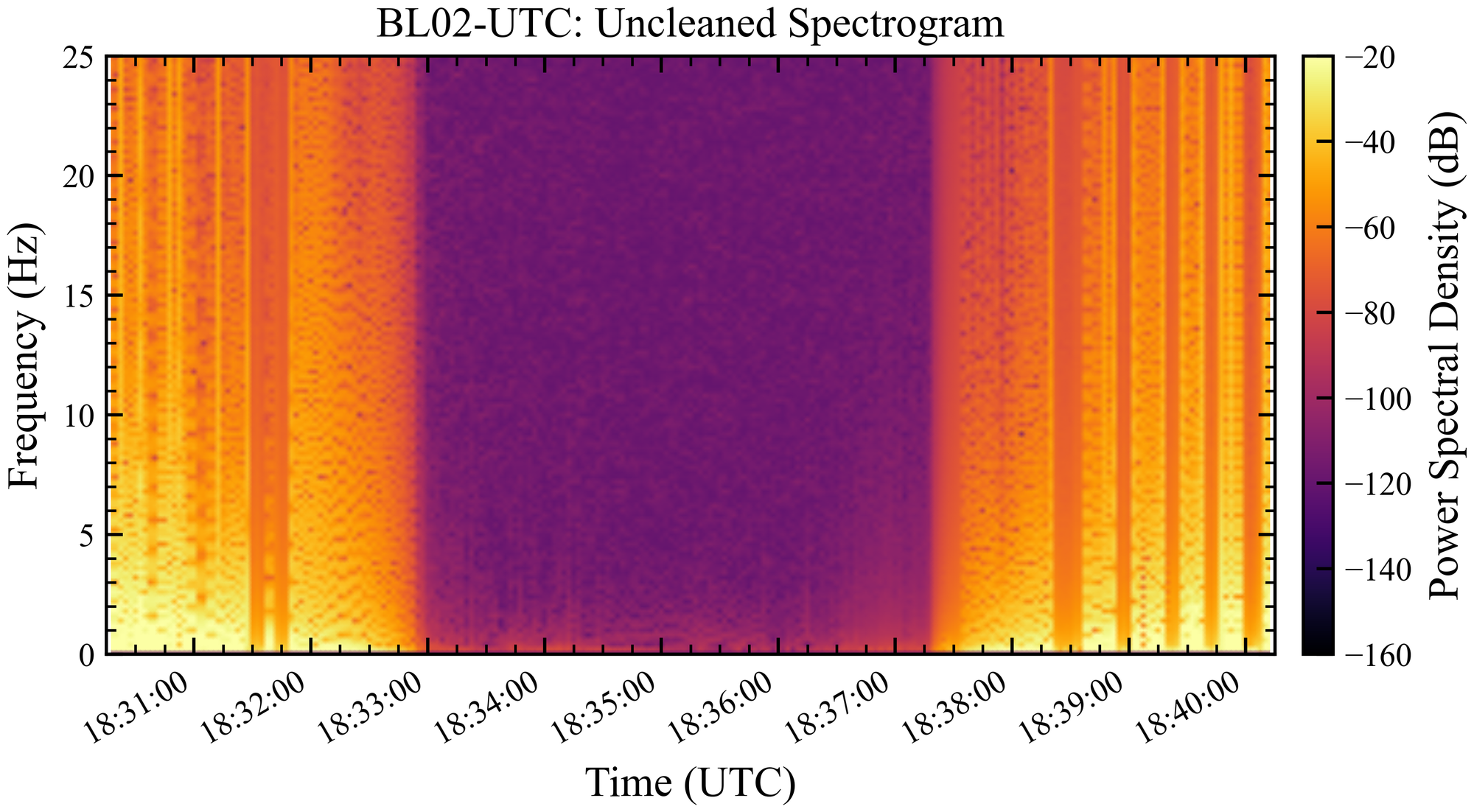}
\par\small \textbf{(b) Raw spectrogram BL02}
\end{minipage}

\vspace{6pt}

\begin{minipage}[t]{\columnwidth}
\centering
\includegraphics[width=0.95\linewidth]{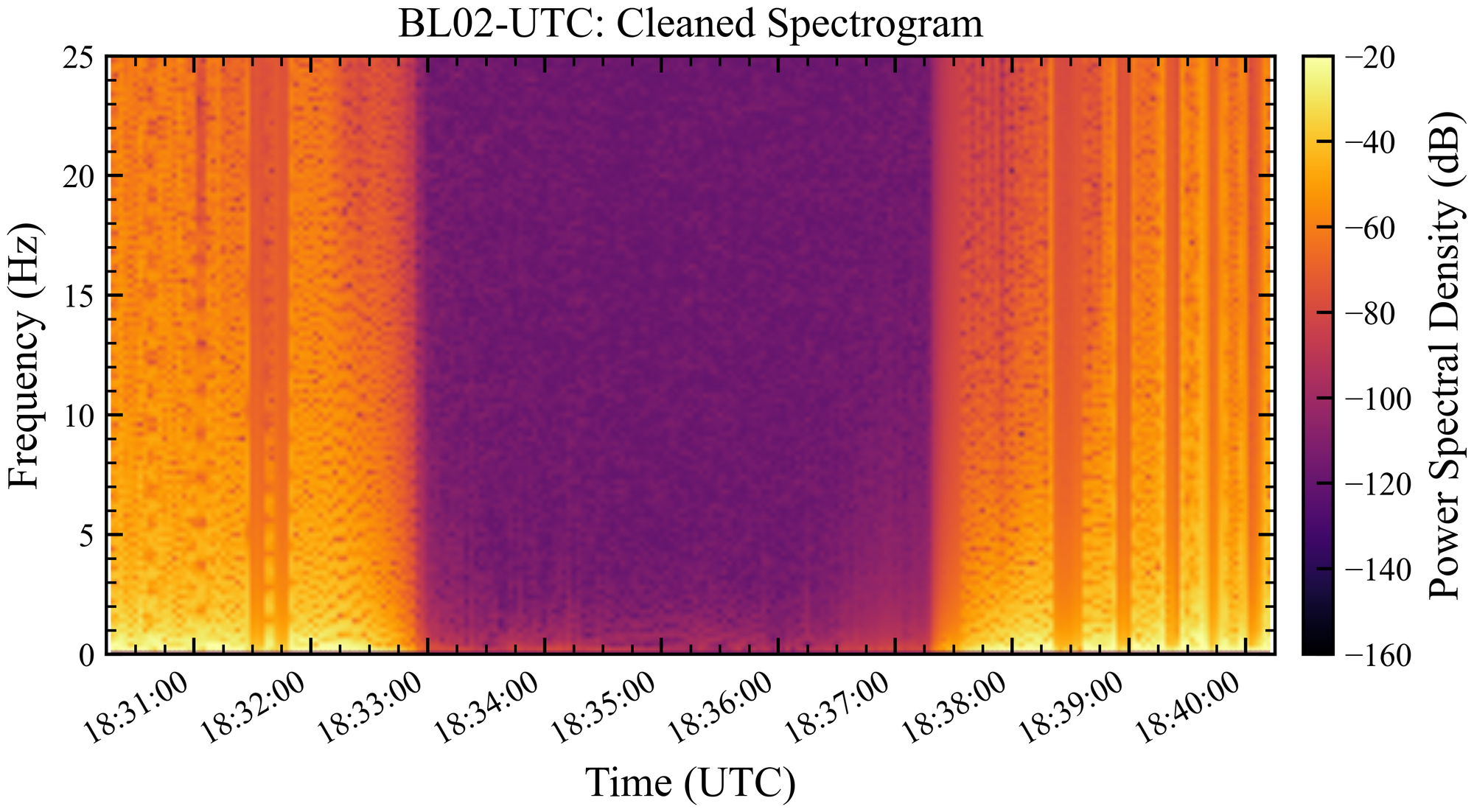}
\par\small \textbf{(c) Cleaned spectrogram BL02}
\end{minipage}

\caption{\textbf{Light-curve cleaning and spectrogram construction for the HAB BL02 photodiode.}
(a) Raw and cleaned voltage time series. (b) Spectrogram before artifact suppression. (c) Spectrogram after artifact suppression. The cleaned BL02 spectrogram shows no persistent narrow-band frequency feature comparable to a shadow band-like signal, consistent with the other HAB photodiode measurements.}
\label{fig:BL02}

\end{figure}

%-----

\begin{figure}

\centering

\begin{minipage}[t]{\columnwidth}
\centering
\includegraphics[width=0.95\linewidth]{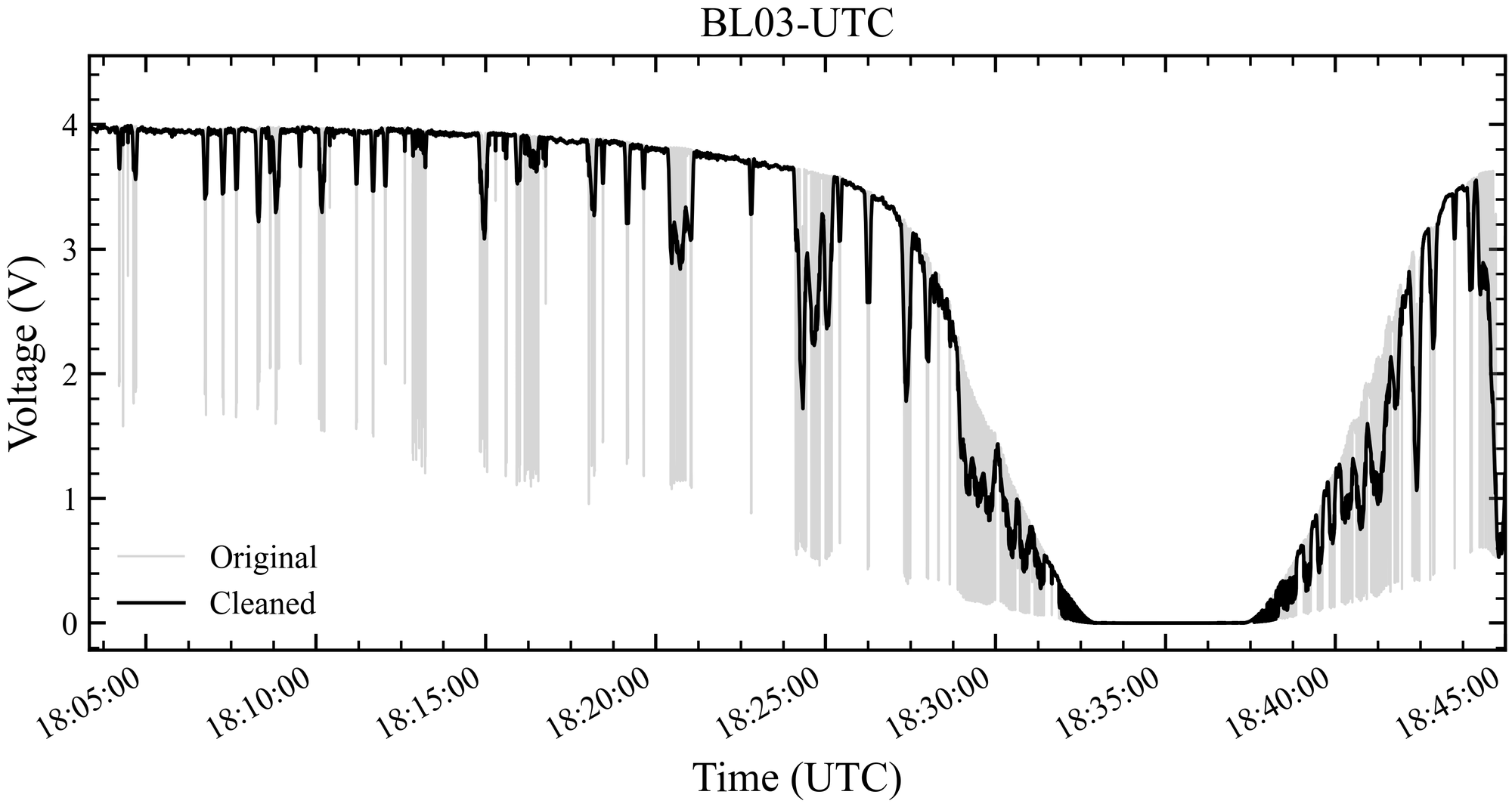}
\par\small \textbf{(a) Light curve BL03}
\end{minipage}

\vspace{6pt}

\begin{minipage}[t]{\columnwidth}
\centering
\includegraphics[width=0.95\linewidth]{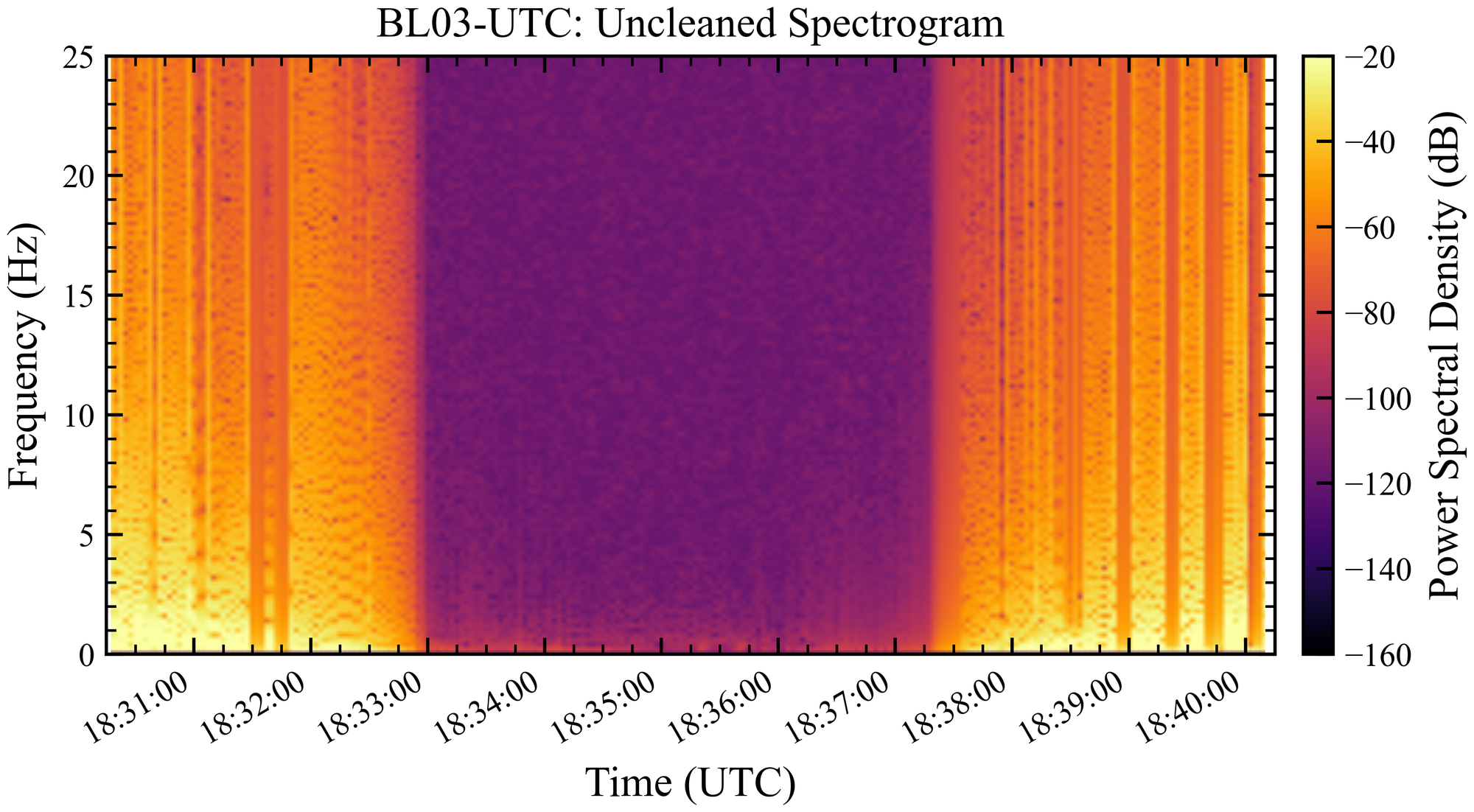}
\par\small \textbf{(b) Raw spectrogram BL03}
\end{minipage}

\vspace{6pt}

\begin{minipage}[t]{\columnwidth}
\centering
\includegraphics[width=0.95\linewidth]{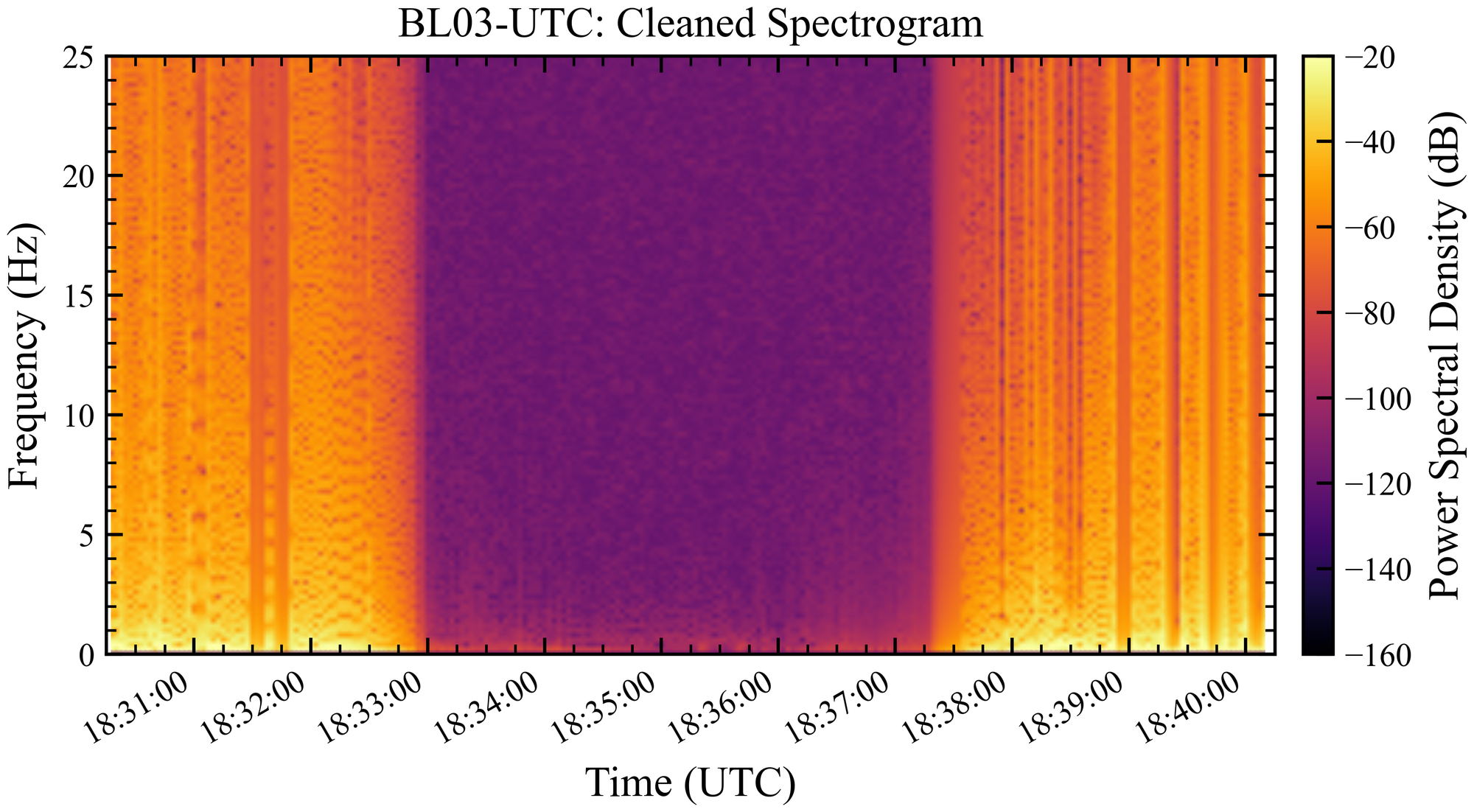}
\par\small \textbf{(c) Cleaned spectrogram BL03}
\end{minipage}

\caption{\textbf{Light-curve cleaning and spectrogram construction for the HAB BL03 photodiode.}
(a) Raw and cleaned voltage time series. (b) Spectrogram before artifact suppression. (c) Spectrogram after artifact suppression. The cleaned BL03 spectrogram shows no persistent narrow-band frequency feature comparable to a shadow band-like signal, consistent with the other HAB photodiode measurements.}
\label{fig:BL03}

\end{figure}

\begin{figure}

\centering

\begin{minipage}[t]{\columnwidth}
\centering
\includegraphics[width=0.95\linewidth]{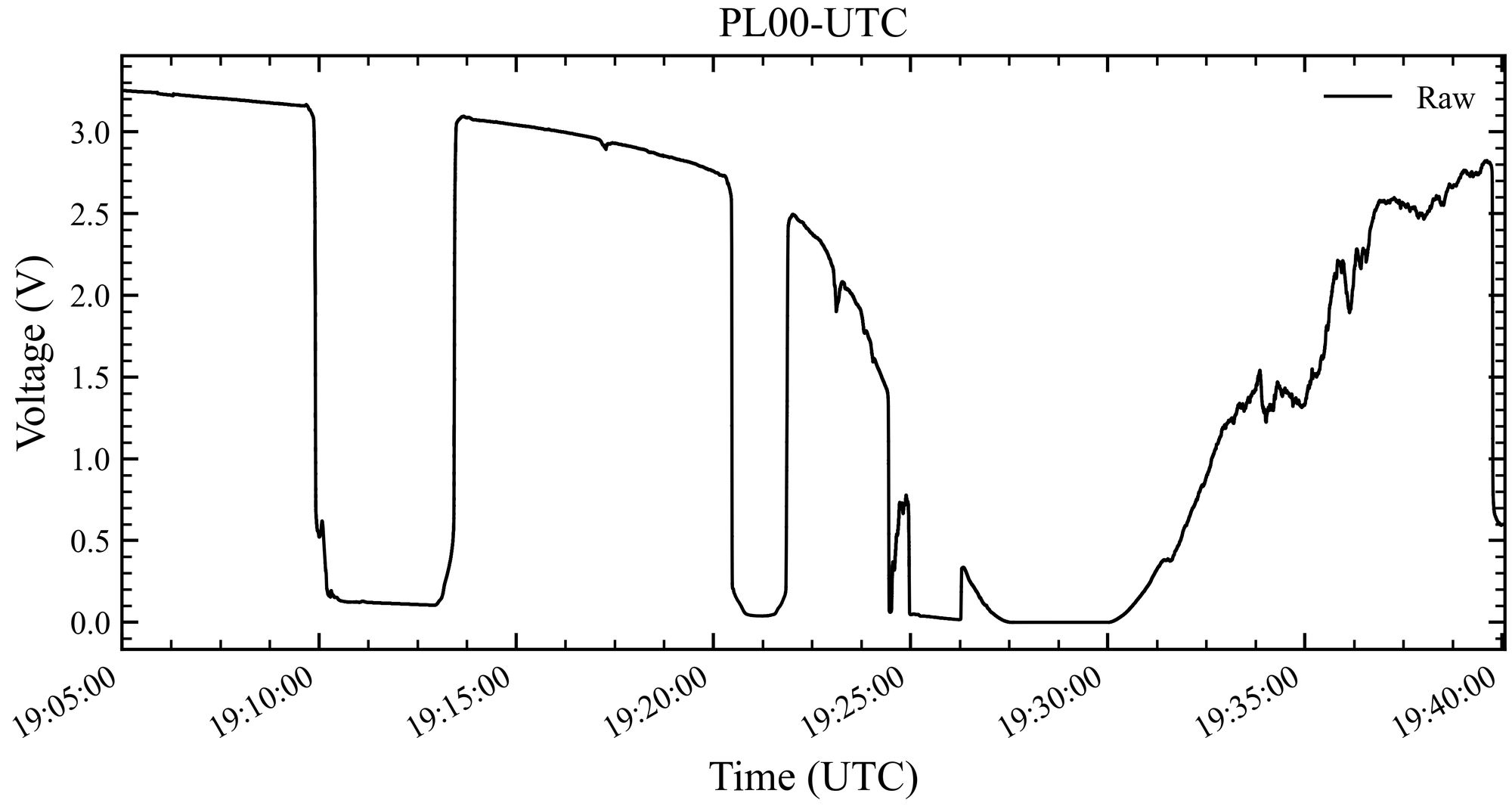}
\par\small \textbf{(a) Light curve PL00}
\end{minipage}

\vspace{6pt}

\begin{minipage}[t]{\columnwidth}
\centering
\includegraphics[width=0.95\linewidth]{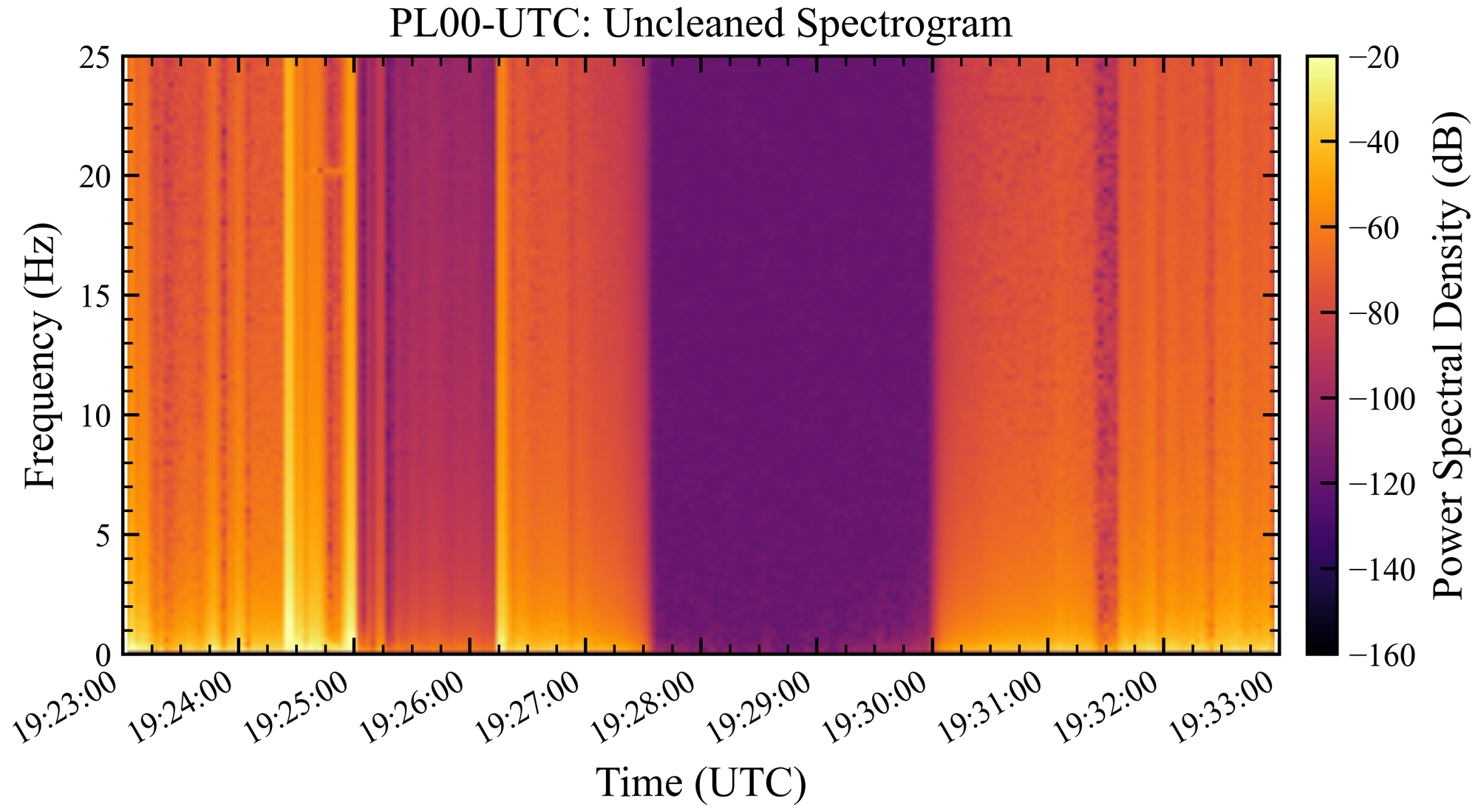}
\par\small \textbf{(b) Raw spectrogram PL00}
\end{minipage}

\caption{\textbf{Light curve and spectrogram for the aircraft-mounted PL00 photodiode.}
(a) Raw aircraft photodiode voltage time series from northeastern Vermont. (b) Spectrogram computed using the same procedure used for the HAB data. Cleaning was not applied because payload-line obstruction artifacts were not present in the aircraft data. No persistent narrow-band feature near 4.5 Hz, or at any other stable frequency, is visible in the aircraft spectrogram.}
\label{fig:PL00}

\end{figure}

\begin{figure}

\centering

\begin{minipage}[t]{\columnwidth}
\centering
\includegraphics[width=0.95\linewidth]{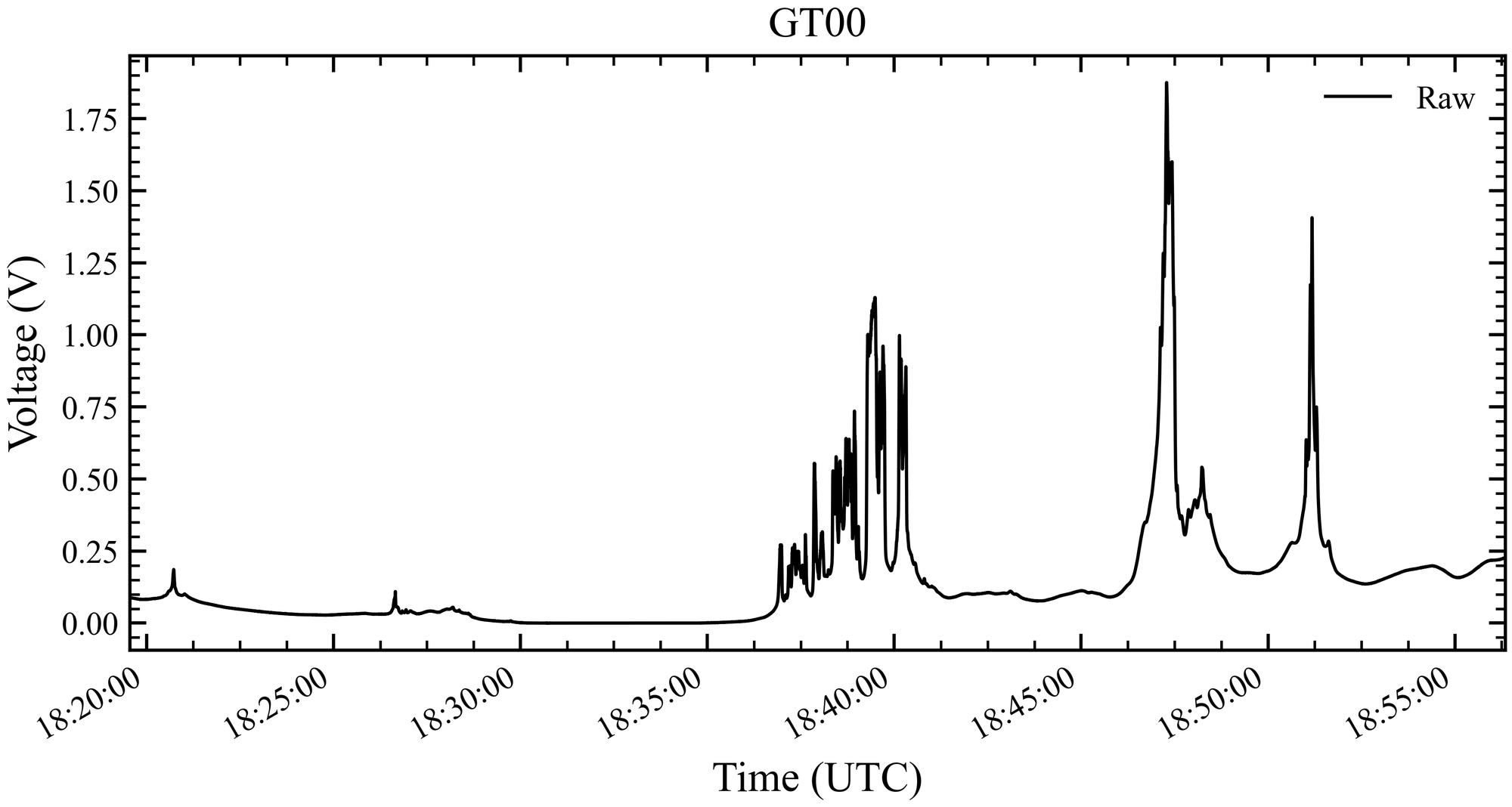}
\par\small \textbf{(a) Light curve GT00}
\end{minipage}

\vspace{6pt}

\begin{minipage}[t]{\columnwidth}
\centering
\includegraphics[width=0.95\linewidth]{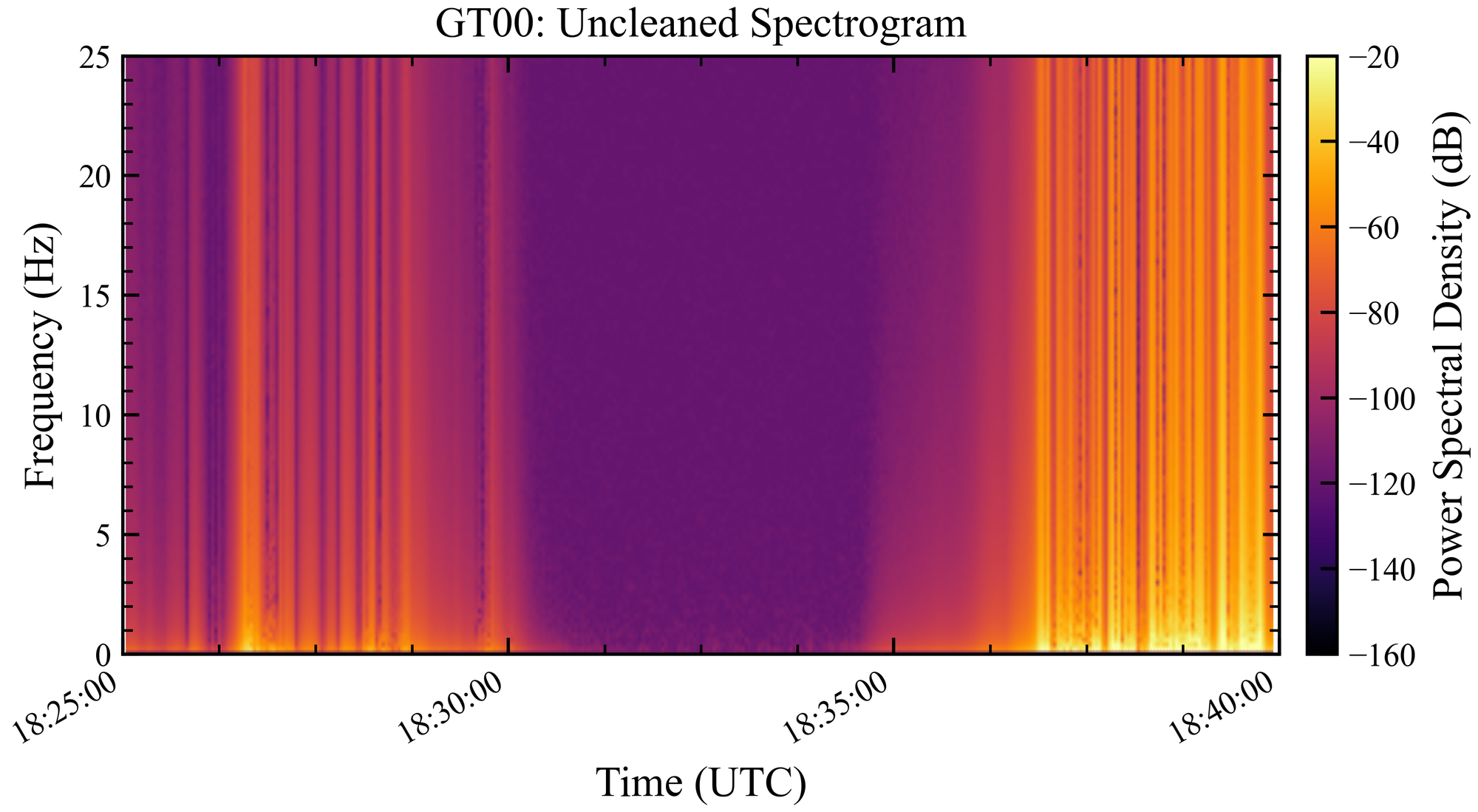}
\par\small \textbf{(b) Raw spectrogram GT00}
\end{minipage}

\caption{\textbf{Light curve and spectrogram for the Concan ground-based GT00 photodiode.}
(a) Raw ground-based photodiode voltage time series from Concan, Texas. (b) Spectrogram computed using the same procedure used for the other photodiode data. Cloud cover during totality strongly affected the ground light curve, limiting the usefulness of these data for determining whether shadow bands were present near the surface at Concan.}
\label{fig:GT00}

\end{figure}
\clearpage
\end{document}